\documentstyle{article}

\newcommand{\phnb}[1]{\Phi_{{#1}}}
\newcommand{\ph}[1]{\mbox{$\phnb{{#1}}$}}
\newcommand{\pht}{\mbox{$\tilde{\Phi}_{10}$}}
\newcommand{\Dmd}{D_{\mu}}
\newcommand{\Dmu}{D^{\mu}}
\newcommand{\sprod}[2]{({#1} \times {#2})_S}

\newcommand{\bequ}{\begin{equation}}
\newcommand{\eequ}{\end{equation}}
\newcommand{\bequa}{\begin{equation} \begin{array}{lll}}
\newcommand{\eequa}{\end{array} \end{equation}}
\newcommand{\Nr}{u}
\newcommand{\tsnb}{T_s}
\newcommand{\ts}{\mbox{$\tsnb$}}
\newcommand{\Rt}{\tilde{R}}
\newcommand{\pzero}[1]{\mbox{$\phi^{(0)}_{{#1}}$}}
\newcommand{\WL}[1]{W_L^{{#1}}}
\newcommand{\WLpm}{\mbox{$\WL{\pm}$}}
\newcommand{\X}[2]{X_{{#1}}^{{#2}}}	
\newcommand{\Xpm}{\mbox{$\X{i}{\pm}$}}
\newcommand{\Y}[2]{Y_{{#1}}^{{#2}}}	
\newcommand{\Ypm}{\mbox{$\Y{i}{\pm}$}}
\newcommand{\WR}[1]{W_R^{{#1}}}
\newcommand{\WRpm}{\mbox{$\WR{\pm}$}}
\newcommand{\lqg}[2]{X_{S{#1}}^{{#2}}}	
\newcommand{\lqgpm}{\mbox{$\lqg{i}{\pm}$}}
\newcommand{\Xg}[2]{{X_{{#1}}'}^{{#2}}}	
\newcommand{\Xgpm}{\mbox{$\Xg{i}{\pm}$}}
\newcommand{\Yg}[2]{{Y_{{#1}}'}^{{#2}}}	
\newcommand{\Ygpm}{\mbox{$\Yg{i}{\pm}$}}
\newcommand{\gen}[1]{\mbox{$\tau^{{#1}}$}}

\newcommand{\ets}{\eta_{2}^{2}}
\newcommand{\rs}{r_s}
\newcommand{\rl}{r_{ew}}
\newcommand{\NO}{Nielsen-Olesen}

\newcommand{\dr}{\partial_r}
\newcommand{\dth}{\partial_{\theta}}
\newcommand{\eith}[1]{e^{i{#1}\theta}}
\newcommand{\emith}[1]{e^{-i{#1}\theta}}
\newcommand{\epmith}[1]{e^{\pm i{#1}\theta}}
\newcommand{\empith}[1]{e^{\mp i{#1}\theta}}

\newcommand{\fl}{\Psi_L}
\newcommand{\flc}{\Psi_L^c}
\newcommand{\flb}{\bar{\Psi}_L}

\newcommand{\fr}{\Psi_R}
\newcommand{\sigt}{i \sigma_2}
\newcommand{\gm}[1]{g_{{#1}}}
\newcommand{\gma}{g'_{2}}
\newcommand{\gmat}{\tilde{g}'_{2}}
\newcommand{\ka}{\kappa}
\newcommand{\kt}{\tilde{\kappa}}

\newcommand{\vd}[1]{\hat{d}_{{#1}}}
\newcommand{\vu}[1]{\hat{u}_{{#1}}}
\newcommand{\ve}{\hat{e}^-}
\newcommand{\vn}{\hat{\nu}}
\newcommand{\vuc}[1]{\hat{u}^c_{{#1}}}
\newcommand{\vdc}[1]{\hat{d}^c_{{#1}}}
\newcommand{\vep}{\hat{e}^+}
\newcommand{\vnc}{\hat{\nu}^c}
\newcommand{\vl}{\hat{\lambda}}
\newcommand{\vv}[1]{\hat{{#1}}}

\newcommand{\vvp}[1]{\hat{{#1}}'}
\newcommand{\vvcp}[1]{\hat{{#1}}'^c}
\newcommand{\eps}[1]{\epsilon_{{#1}}}

\newcommand{\Hg}[1]{H^{{#1}}}
\newcommand{\Hgt}[1]{\tilde{H}^{{#1}}}
\newcommand{\Hgg}{{\cal H}^0}

\newcommand{\pU}{\psi_U}
\newcommand{\pL}{\psi_D}
\newcommand{\UU}[1]{{#1}_U}
\newcommand{\LL}[1]{{#1}_D}
\newcommand{\UL}[1]{{#1}_{U,D}}

\newcommand{\Vs}{\frac{B'}{\sqrt{6}}}

\newcommand{\pharrow}[1]{\stackrel
	{ \mbox{\scriptsize $\phnb{{#1}}$} }{\longrightarrow}}
\newcommand{\spa}{\hspace{.1in}}

\begin{document}

\title{\bf Microphysics of $SO(10)$ Cosmic Strings} 
\author{Anne-Christine Davis and Stephen C. Davis \\ \\
\em Department of Applied Mathematics and Theoretical Physics,\\
\em University of Cambridge, Cambridge, CB3 9EW, UK}

\maketitle

\begin{abstract}
We uncover a rich microphysical structure for $SO(10)$ cosmic strings.
For the abelian string the electroweak symmetry is restored around
it in a region depending on the electroweak scale. Four distinct
nonabelian strings are found. Some of these also
restore the electroweak symmetry. We investigate the zero mode
structure of our strings. Whilst there are right handed neutrino zero modes 
for the abelian string, they do not survive the electroweak phase
transition in the case of the lowest energy solution. We elucidate the
zero mode structure for the nonabelian strings above and below the
electroweak phase transition. We
consider the generalisation of our results to other theories and
consider the cosmological consequences of them. 
\end{abstract}

\vfill

\begin{flushright} DAMTP/96-72 \end{flushright}

\vfill

\newpage

\section{Introduction}
Topological defects arise as a result of phase transitions in the
early universe \cite{strings}. Such topological defects, in particular
cosmic strings, resulting from a grand unified theory (GUT),
generate density perturbations which could explain the observed
large-scale structure and anisotropy in the microwave background, see
\cite{strings} and the references therein. They also provide an
important window into the physics of the very early universe. For
example, in the core of the cosmic strings the underlying GUT symmetry
is restored, resulting in baryon violating processes being
unsuppressed. This can catalyse proton decay \cite{proton}, and the
decay of string loops can explain the observed baryon asymmetry
\cite{asymmetry}.

In recent years it has become increasingly apparent that cosmic
strings have a richer microstructure than previously thought
\cite{Warren,Goodband}. In particular, at subsequent phase transitions
the core of the cosmic string acquires additional features. For
example, the string can cause electroweak symmetry restoration in a
much larger region around it, proportional to the electroweak scale
itself \cite{Warren, Goodband}. This new microphysical structure has
been used to provide a new scenario for electroweak baryogenesis
\cite{baryogenesis}, and to investigate the current-carrying
properties of cosmic strings \cite{currenta,currentb}.

Previous work considered the simplest extension to the Standard
Model that would allow the formation of strings. A $U(1)$ symmetry,
whose breaking produced an abelian string, was added to the usual
Standard Model symmetries. The resulting theory had two coupling
constants, of arbitrary ratio. It was shown that if the ratio was
large enough, the electroweak Higgs field would not only be zero at
its centre, but would also wind like a
string. Whether this is likely to happen with phenomological strings
can be found by considering a realistic grand unified theory, where
there is less arbitrariness. 

By using a larger gauge group it is also possible to consider the effects
of nonabelian strings, which could not occur in the theories
considered in \cite{Warren,Goodband}. Nonabelian strings have
significantly different behaviour to abelian strings, since the
associated string generators do not all commute with the Standard
Model fields, or the other gauge fields. It is thus necessary to approach
them in a slightly different way.

As well as symmetry restoration, the presence of a string may allow
the formation of non-trivial zero energy fermion solutions, as
discussed by Jackiw and Rossi \cite{Jackiw}, for an abelian string. If
such zero modes exist, a superconducting current
can flow along the string which may have long range effects
\cite{Witten}. It is possible that the electroweak Higgs
field, because of its string-like form, could also allow the formation
of such zero modes \cite{currentb,Stern,Witten}.     

In this paper we examine these issues in detail for strings formed in
a realistic grand unified theory based on $SO(10)$.
In section \ref{SOtenGUT} the grand unified theory to be used is
outlined. The possible strings that form in it prior to the
electroweak phase transition are discussed in section
\ref{GUTstrings}. In section \ref{EWSymBrk} the effect of the strings
on the electroweak symmetry are considered, and the approximate form of the
fields, as well as an estimate of the energy is found. In section
\ref{OtherSymmRest} some other, simpler symmetry restorations occuring
in the theory are discussed, in particular that of the intermediate $SU(5)$
symmetry. In section \ref{FermionZM} the possible existence of fermion zero
modes for abelian and nonabelian strings is investigated, including
cases in which two Higgs fields effect the fermion fields. Although
one specific theory is considered, many of the results generalise to
other theories. The implications of our results for such theories are
discussed in section \ref{OtherGUTs}. In section \ref{SOtenConc} we
summarise our results and discuss the conclusions.

\section{An $SO(10)$ Grand Unified Theory}
\label{SOtenGUT}

A realistic GUT which has a symmetry breaking pattern which produces
strings is $SO(10)$. Its properties have a reasonable agreement
with physical results. Consider the symmetry breaking
\bequa
SO(10) & \pharrow{126} & SU(5) \times Z_2 \\ 
& \pharrow{45} & SU(3)_c \times SU(2)_L \times U(1)_Y \times Z_2 \\
& \pharrow{10} & SU(3)_c \times U(1)_Q \times Z_2 \label{symbreak}
\eequa
where \ph{N} transforms under the {\bf N} representation of $SO(10)$.
The actual grand unified gauge group is $Spin(10)$, the covering group of
$SO(10)$, but for simplicity the symmetry breaking is shown in
terms of the Lie algebras. 
The discrete $Z_2$ symmetry formed by the \ph{126} Higgs field
leads to the formation of a variety of cosmic strings.  Comparison
of the effects of the various symmetry breakings is simplified by
expressing everything in terms of the same representation. Since
${\bf 126} + {\bf 10} = \sprod{\bf 16}{\bf 16}$ this is possible. 
Conveniently, {\bf 16} is also the representation that acts on the
fermions. The $SO(10)$ fermions consist of the usual standard model
fermions, plus a right handed neutrino. The fermionic part of the
theory is then expressed in terms of the left-handed fermions and the
charge conjugates of the right-handed fermions.

The maximal subgroup of $SO(10)$ is actually $SU(5) \times U(1)_P$, and 
$P$ can be used to decompose $SO(10)$ into representations of $SU(5)$
\bequ
 {\bf 16} \longrightarrow {\bf 1}_5 + {\bf 10}_1 + {\bf \bar{5}}_{-3}
\eequ
where the subscripts are the eigenvalues of $P$. ${\bf 126}$ and ${\bf
10}$ can be similarly decomposed by considering symmetric products of
{\bf 16}. 
\bequ
{\bf 126} \longrightarrow {\bf 1}_{10} + \ldots \hspace{.7in}
{\bf 10} \longrightarrow {\bf 5}_{-2} + {\bf \bar{5}}_2
\eequ
$P$ can also be used to describe the non-trivial element of
the discrete symmetry of (\ref{symbreak}), which is $D = \exp(2\pi i P/10)$.

The vacuum expectation value of \ph{126} has a magnitude of
$\frac{\eta_{1}}{\sqrt{2}}$, which is of order $10^{15}$ GeV. It is in the
${\bf 1}_{10}$ component of {\bf 126}, and so must be equal to
$\frac{\eta_{1}}{\sqrt{2}}(\Nr \times \Nr)$, where $\Nr$ is in the ${\bf
1}_5$ component of the {\bf 16} representation of $SU(5)$ (the
corresponding field in the fermion representation is the charge conjugate
of the right-handed neutrino). In $SU(5)$ the VEV of the equivalent of
$\ph{10}$ is in the chargeless component of ${\bf 5}_{-2}$. This choice
can create problems in $SO(10)$ (see section \ref{LTWstrNnZM}), so the VEV
of \ph{10} is made up of both the chargeless components of {\bf 10}
instead.  If $\Hg{0}$ and $\Hgt{0}$ are the chargeless components of 
${\bf 5}_{-2}$ and ${\bf \bar{5}}_2$ respectively, \ph{10} can be
expressed as $\frac{\eta_{2}}{\sqrt{2}}\Hgg$, where 
$\Hgg = \Hg{0}\ka + \Hgt{0}\kt$. If the fermion field phases are defined
appropriately, $\ka$ and $\kt$ will be real and positive, and
$\ka^{2} + \kt^{2} = 1$. Since {\bf 10} is contained in $\sprod{\bf
16}{\bf 16}$, $\Hgg$ can be expressed as a sum of symmetric products
of components of {\bf 16}s. The {\bf 45} is contained in ${\bf 16}
\times {\bf \bar{16}}$. \ph{45}'s effect on the formation of strings
is far less significant than the other Higgs fields, so it can be
ignored for now.
 
Strings can form at the first $SO(10) \longrightarrow SU(5) \times Z_2$
symmetry breaking. In this case \ph{126} is not constant, and takes the form
$e^{i \theta \tsnb}\pzero{126}(r)$. $\pzero{126}$ is independent
of $\theta$, and satisfies the boundary condition 
$\pzero{126}(\infty) = \frac{\eta_{1}}{\sqrt{2}}\Nr \times \Nr$. \ts\ is
made up of the broken generators of $SO(10)$, and must give a
single-valued \ph{126}.  
If $e^{2 \pi i \tsnb} = U \times D$ (where $U$ is in $SU(5)$), then while 
\ph{126} will be single valued, it will not be topologically equivalent to
$\ph{126} =$ constant, and so the string will be topologically
stable. If $e^{2 \pi i \tsnb} = U \times I$ the string is not
topologically stable, but may have a very
long lifetime, and so still be physically significant \cite{Witten}. 

The Lagrangian of the system is 
\bequ \begin{array}{r@{}l}
{\cal L} = (\Dmd \ph{126})^{\ast}&(\Dmu \ph{126})
 +(\Dmd \ph{10})^{\ast}(\Dmu \ph{10})
 + (\Dmd \ph{45})^{\ast}(\Dmu \ph{45})\\
 &{}- \frac{1}{4} F^a_{\mu \nu}F^{\mu \nu a} 
- V(\ph{126},\ph{45},\ph{10})+ {\cal L}_{\mbox{\scriptsize fermions}} 
\label{lagrangian}
\eequa
where $\Dmd = \partial_{\mu} - \frac{1}{2}igA_{\mu}$ and $F_{\mu \nu} =
\partial_{\mu}A_{\nu} - \partial_{\nu}A_{\mu} - 
\frac{1}{2}ig[A_{\mu},A_{\nu}]$. 
There are 45 gauge fields in all, most of which acquire superheavy
masses and so are not observed at everyday temperatures. They consist
of the usual standard model fields, with the $W$-bosons
denoted by $\WL{i}$; $\WR{i}$, which are right-handed versions of the
$\WL{i}$, coupling right handed neutrinos to electrons; some
leptoquark bosons: \Ypm, \Xpm, \lqgpm, where \Ypm\ and \Xpm\ are $SU(5)$
gauge fields;  
some more general gauge fields \Xgpm\ and \Ygpm,
which couple quarks to leptons and different coloured quarks;
 and a fifth diagonal field, $B'$.
The index $i$ takes the values 1,2,3, and is related to colour. Two uncharged
diagonal fields, $S$ and $B$, are
made up of orthogonal linear combinations of $\WR{3}$ and
$B'$. Linear combinations of $B$ and $\WL{3}$ produce the $Z$ boson and
the photon, $A$.
At the first symmetry breaking $S$, \WRpm, \Xgpm, \Ygpm, and \lqgpm\ are
all given superheavy masses. The second stage gives high masses to \Xpm, \Ypm,
and additional masses to \WRpm, \Xgpm, \Ygpm\ and \lqgpm. Finally the third
symmetry breaking gives masses to \WLpm and $Z$, with further
masses being given to the $S$, \WRpm, \Ypm\ and \Ygpm\ fields.

\section{GUT Strings}
\label{GUTstrings}

Neglecting fermions, the Euler-Lagrange equations obtained from
(\ref{lagrangian}) are
\bequ
\Dmd \Dmu \phnb{i} = - \frac{\partial V}{\partial \phnb{i}^{\ast}}
						\label{phiequ}
\eequ
\bequ
\partial_{\mu} F^{\mu \nu a} -
\frac{1}{2}ig{f^{c}}_{ab}A^b_{\mu}F^{\mu \nu c} = -g\mbox{Im}
\sum_i (D^{\nu}\phnb{i})^{\ast}(\tau^a \phnb{i}) \label{gaugeequ}
\eequ
At high temperatures $V$ is such that \ph{126} is the only non-zero
Higgs field. (\ref{phiequ}) and (\ref{gaugeequ}) have various string
solutions. The different solutions correspond to different choices of \ts. 
In the $SO(10) \longrightarrow SU(5) \times Z_2$ symmetry breaking,
21 of $SO(10)$'s 45 generators are broken, and \ts\ is a linear
combination of them. One of them, $P$,  
corresponds to the $U(1)$ symmetry not embedded in $SU(5)$. The
corresponding string is abelian, and has the solution
\bequa
\ph{126} &=& f(r)e^{in\theta \tsnb}\pzero{126}(\infty) \\
A_{\theta} &=& n \frac{2a(r)}{gr} \ts \hspace{1in}
A_{\mu} = 0 \mbox{ \ otherwise} \label{abansatz}
\eequa
where \ts, the string generator, equals $\frac{P}{10}$, and $n$ is
an integer. The non-zero gauge field is required to give a zero
covariant derivative, and hence zero energy, at infinity. It
corresponds to a non-zero $S$ field.
(\ref{abansatz}) can be simplified using $P\Nr = 5\Nr$, to give
$e^{in\theta \tsnb}\pzero{126} = e^{in\theta}\pzero{126}$.
Substituting (\ref{abansatz}) into (\ref{phiequ}) and (\ref{gaugeequ})
gives the \NO\ vortex equations, as would be expected. Regularity at
the centre of the string, and finite energy due to a vanishing covariant
derivative and potential at infinity, imply the boundary conditions
$f(0)=a(0)=0$ and $f(\infty)=a(\infty)=1$.

The situation for the other generators is more complicated. For a
general string generator \ts, the left and right hand sides of (\ref{phiequ})
are proportional to $\tsnb^{2}\pzero{126}$ and \pzero{126} respectively,
which in general are not proportional \cite{Ma}. Thus the solution
(\ref{abansatz}) will not work. This is resolved by expressing 
$\pzero{126}(\infty)$ in terms of the eigenstates of $\tsnb^{2}$, to give 
$\pzero{126}(\infty) = \sum_m \phi_m$, 
where $\tsnb^{2} \phi_m = m^2 \phi_m$. Since \ts\ is Hermitian, $m^2$
will be positive and real. A suitable string solution can now be
constructed
\bequa
\ph{126} &=& e^{i\theta \tsnb}\sum_m f_m(r)\phi_m\\ 
A_{\theta} &=& \frac{2a(r)}{gr} \tsnb \hspace{1in}
A_{\mu} = 0 \mbox{ \ otherwise} \label{nabansatz}
\eequa
In order for \ph{126} to be single valued the various $m$ must all be
integers, and $\tsnb\phi_{0}$ must be zero. The boundary conditions on
$a$ and $f_m$ will be the same as those for (\ref{abansatz}),
except that $f_0$ need not be zero at $r=0$. The simplest examples
of such solutions occur when $\tsnb^{2}\Nr=\frac{n^2}{4}\Nr$ (where $n$
is an integer), in which case
\bequ
\phi_n = \frac{\eta_1}{\sqrt{2}}  \left( 
\frac{1}{2}\Nr \times \Nr + \frac{2}{n^2}\tsnb \Nr \times \tsnb \Nr \right)
 \mbox{ and } \phi_0 = \frac{\eta_1}{\sqrt{2}}  \left( 
\frac{1}{2}\Nr \times \Nr - \frac{2}{n^2} \tsnb \Nr \times \tsnb \Nr \right)
\label{nabphis}\eequ
$\tsnb \Nr$ and $\Nr$ are orthogonal, so $\phi_0$ and $\phi_n$ are
orthogonal. In this case only part of the Higgs
field winds around the string. This type of string was first suggested
by Aryal and Everett \cite{AryalE}, and has been examined in detail by
Ma \cite{Ma}. It turns out to have lower (about half as much) energy
than the abelian string (\ref{abansatz}). This is because the Higgs
field is not forced to be zero at the string's centre, which reduces
the contribution to the energy from the potential terms. Also since
\ph{126} varies less, the covariant derivative terms are smaller.

Of course, such vortex-like solutions are only topological strings if
$e^{2\pi i\tsnb}$ is not contained in $SU(5)$. If $n$ is even this is
not the case, and the solution is topologically equivalent to the
vacuum.  Similarly, odd values of $n$ are all topologically equivalent
to each other, so there is only one topologically distinct type of
string of this form. Strings with higher $n$ can unwind into
strings with lower $n$. The same is true of the abelian string.
However it is possible that the lifetime of an $n>1$ string will be
very long, so in a general theory all values of $n$ should be considered. 
Putting $\tsnb \longrightarrow n\tsnb$ in (\ref{nabansatz}) and $n=1$ in
(\ref{nabphis}) gives a similar ansatz to (\ref{abansatz}), making
comparison easier.

As shown in \cite{Ma} the most general potential reduces to a
different form to that of the abelian case, and leads to these
equations for $a$ and the $f_{m}$'s   
\bequ 
f_0'' + \frac{f_0'}{r} = \eta_1^2 \left[\lambda_1
\left(\frac{f_1^2 + f_0^2}{2} - 1\right) - \mu_1
\left(\frac{f_1^2 - f_0^2}{2}\right)\right]f_0
\label{nabequa} \eequ \bequ
f_1'' + \frac{f_1'}{r} - n^2\frac{(1-a)^2}{r^2}f_1 = 
\eta_1^2 \left[\lambda_1\left(\frac{f_1^2 + f_0^2}{2} -
1\right) + \mu_1 \left(\frac{f_1^2 - f_0^2}{2}\right)\right]f_1
\label{nabequb} \eequ \bequ
a'' - \frac{a'}{r} = -\frac{1}{2}g^2 \eta_1^2 (1-a)f_1^2 
\label{nabequc} \eequ
where $\mu_1$ and $\lambda_1$ are such that $f_1 (\infty)$ and 
$f_0 (\infty)$ will both be 1. The corresponding equations for the
abelian string ($\ts = P/10$) are
\bequ
f'' + \frac{f'}{r} - n^2 \frac{(1-a)^2}{r^2}f =
				 \eta_1^2 \lambda_1 (f^2 - 1)f
\label{abequa} \eequ \bequ
a'' - \frac{a'}{r} = -\frac{5}{2}g^2 \eta_1^2 (1-a)f^2 
\label{abequb} \eequ
The above nonabelian strings are in fact all $SU(2)$ strings. There
are other more complicated possibilities, for which $\tsnb^2 \Nr$ is not
proportional to $\Nr$, but none of these are topologically stable. We
shall only consider topologically stable strings, and the closely
related solutions with higher winding numbers.
If a single valued charge operator is also required there are just a few 
possibilities for \ts, which can be classified in terms of the non-zero
gauge fields around the string. These are all equivalent under
$SU(5)$, but not under $SU(3)_c \times U(1)_Q$, so they will be
distinct after the electroweak symmetry breaking. Apart from the
abelian string the four cases correspond to non-zero \WRpm, \lqgpm,
\Xgpm\ and \Ygpm\ fields. Under $SU(3)_c \times U(1)_Q$
any linear combination of \lqgpm\ generators can be gauge transformed
into any other combination, thus they are equivalent. The same is true
for the other generators, so there are just 5 distinct types
of string at low temperatures. The 4 nonabelian strings can be labelled
by their gauge fields. Under $U(1)_Q$, nonabelian strings
with winding number $-n$ are gauge equivalent to ones with winding
number $n$ (for any choice of \ts), so it is sufficient to
consider only $n > 0$ strings. 

Since all the nonabelian strings are gauge equivalent under $SU(5)$,
they have the same energy, which is about half that of the abelian
string \cite{Ma}. Although these strings give single valued electric
charge, they do not all give single valued colour charge, and so some
are Alice strings. This does not mean that they are unphysical, as
discussed in \cite{suthree}.  

\section{The Electroweak Symmetry Breaking} 
\label{EWSymBrk}

Topological strings only form in a symmetry breaking $G \longrightarrow H$
if $\pi_1 (G/H) \neq I$. This is not the case at the third (\ph{10})
symmetry breaking, so such strings do not form there. It is still possible
for \ph{10} to wind, and for string-like solutions to appear \cite{Vachaspati}.
However, since it is energetically favourable for the Higgs field to
unwind, they are not completely stable (although they could take a
long time to decay). 

The situation is different in the presence of a topological string,
formed at a previous symmetry breaking.
If the electroweak Higgs field took its usual
constant vacuum expectation value now, and the gauge fields from the
string did not annihilate it, its covariant derivative would be
non-zero everywhere. At large $r$ it would be proportional to $1/r$,
and so it would have logarithmically divergent energy. This can be avoided by
allowing \ph{10} to wind like the GUT string. In order for it to be
single valued, it may be necessary to alter the string generator,
\ts. Alternatively it may be possible cancel the effect of the string
gauge fields by adding other gauge fields which have an opposite
effect on \ph{10}. The most energetically favourable solution is
likely to be a combination of these two alterations. Unlike the
electroweak strings considered by Vachaspati \cite{Vachaspati}, the
resulting string-like solutions would be stable. 

\subsection{The Abelian $U(1)$ String}
 
For the abelian string, the covariant derivative can be made to vanish
by letting \ph{10} wind. But $\ts \Hg{0} = -\frac{1}{5}\Hg{0}$, and 
$\tsnb \Hgt{0} = \frac{1}{5}\Hgt{0}$, so
$e^{2\pi i n\tsnb}\pzero{10} \neq \pzero{10}$ in
general. When $n$ is not a multiple of 5, just allowing
\ph{10} to wind is insufficient, since this will give a multi-valued
\ph{10}. Instead \ts\ must be replaced by another generator, $\tilde{\tsnb}$,
which will give a single valued \ph{10}. The gauge field must also be
changed to give a vanishing covariant derivative. This
alteration must not affect the GUT string, as any change would give a
large increase in energy. Thus $\tilde{\tsnb}$ needs to be of the form
$\tsnb + R$, where $R$ annihilates $\Nr$ and  $\ts \Nr$, but has some
affect on $\Hg{0}$. If $R\Hg{0} = k\Hg{0}$ and $R\Hgt{0} = -k\Hgt{0}$, 
($k$ is real since $R$ is Hermitian), a suitable ansatz is
\bequa
\ph{10} &=& \frac{\eta_2}{\sqrt{2}}h(r)e^{in(\tsnb + R)\theta}
\Hgg = \frac{\eta_2}{\sqrt{2}}h(r)
\left(\eith{m}\Hg{0}\ka +\emith{m}\Hgt{0}\kt\right) \\
A_{\theta} &=& n\frac{2}{g}\left(\frac{a(r)}{r}\tsnb + \frac{b(r)}{r} R \right)
\hspace{1in} A_{\mu} = 0 \mbox{ \ otherwise} \label{EWabansatz}
\eequa
where $m = n\left(k-\frac{1}{5}\right)$, and $a(r)$ is determined by
the GUT string. Now the covariant derivative of \ph{10} does vanish at
infinity (provided $b(\infty)=1$), and if $k$ is such that $m$ is an
integer, \ph{10} will still be single-valued. Such an $R$ does exist
in $SO(10)$, and it is proportional to the generator of $Z$. Strictly,
the variation of \ph{10} will effect the GUT string, but the effect is of order
$\ets/\eta_1^2 \sim 10^{-26}$, and so can be ignored. (\ph{10}
takes the role of the Weinberg-Salam Higgs field and so its usual
vacuum expectation value, $\frac{\eta_2}{\sqrt{2}}$, is of order
$10^2$ GeV.) The functions $h(r)$ and $b(r)$ obey the same boundary
conditions as $f(r)$ and $a(r)$ in (\ref{abequa},\ref{abequb}), for the same
reasons, except when $m=0$, in which case $h(0)$ need not be zero. 
This gives the field equations
\bequ
h'' + \frac{h'}{r} - n^2\frac{\left[k(1-b)
- \frac{1}{5}(1-a)\right]^2}{r^2} h = \ets \lambda_2 (h^2 - 1)h
\label{abEWphi} \eequ
\bequ
b'' - \frac{b'}{r} = -\frac{2}{5}g^2 \ets
\left\{(1-b)-\frac{1}{5k}(1-a)\right\}h^2
\label{abEWgauge}
\eequ
Strictly they should include contributions arising from
$|\ph{126}|^2|\ph{10}|^2$ terms in the Lagrangian. However, outside
the string core $|\ph{126}|$ is constant, and the cross terms can be
absorbed into the constant potential terms. Inside the string this is
not so, but the effect of the gauge terms on \ph{10} far outweighs
that of the potential terms, as noted in \cite{Goodband}. Thus
$|\ph{126}|$ can be taken as being constant everywhere.
These are similar in form to the \NO\ vortex equations,
except for the extra gauge term. They are also similar in that they cannot
generally be analytically solved. However, using a similar approach to
\cite{Warren} it is possible to obtain an approximate trial solution,
and use it to estimate the energy of (\ref{EWabansatz}).

Defining $\rs$ to be the radius of the string (outside of which
$|\ph{126}|$ takes its usual VEV), and $\rl$ to be the radius of the
region in which $|\ph{10}|$ does not take its usual VEV, the solution can be 
approximated separately in three regions.  If $g$, $\lambda_1$ and
$\lambda_2$ are of order 1, $\rs$ will be of order $\eta_1^{-1}$
and $\rl$ will be of order $\eta_2^{-1}$. At short distances the
potential, and additional gauge term, $b(r)$, can be neglected in
(\ref{abEWphi}), as the other terms dominate.  Inside the string
itself, ($r < \rs$) $a(r) \sim r^2$ can also be ignored giving $h(r)
= C r^{|m|}$ (the other solution is not used since $h(0)$ must be
zero). $b(r)$ is made up of two terms, an $r^{2|m|+2}$ term, proportional
to $C$, and an arbitrary $r^2$ term. For $\rs < r < \rl$, $a(r)$ is
approximately 1, and so $h = A r^{|nk|} + B r^{-|nk|}$. Requiring
continuity of $h$ and $h'$ at $r = \rs$, ensures that $B \sim
A\rs^{2|nk|}$, and so it can be neglected away from the string. The situation
is slightly different for $m=0$. In this case $h(r) = C$ for $r<\rs$
(the second, logarithmic solution is not allowed by regularity at
$r=0$ \cite{Ma}) and $h(0) \neq 0$. The solution will still take the
above form for $r>\rs$. Matching the solutions gives $B = A\rs^{2|nk|}$,
and $C=2A\rs^{|nk|}$, thus $B$ can be neglected as before.
Inserting this into (\ref{abEWgauge})
for $\rs < r < \rl$ gives $b(r) = F r^{2(|nk|+1)} + G r^2$, where $G$
is arbitrary and $F$ is related to $A$. Matching $h$ and $b$ at
$r=\rl$ reveals that to first order $F$ can be neglected, and so a
sensible trial solution is
\bequ
h(r) = \left\{ \begin{array}{c}
 \left(\frac{r}{\rl}\right)^{|nk|} \\  1 \end{array} \right. \hspace{.5in}
b(r) = \left\{ \begin{array}{cl}
 \left(\frac{r}{\rl}\right)^2 & r < \rl \\
		  1 & r > \rl \end{array} \right. \label{abtrial}
\eequ
For $m=0$ this determines $C$ to be $2(\rs/\rl)^{|nk|}$, which
although not equal to zero, is very close.
An estimate of the energy of this string-like solution can now be found by
substituting the trial solution into the Lagrangian. The contribution
from the region $r<\rs$ is suppressed by powers of $\rs$ and can
be neglected. All the contributions are zero for $r > \rl$, thus
\bequ \begin{array}{lll@{}l}
\mbox{Energy} &=& \multicolumn{2}{l}{ 
2\pi\int_{\rs}^{\infty} r dr \left\{
  \left|\dr \ph{10}\right|^2 + \left|D_{\theta}\ph{10}\right|^2
+ \frac{1}{2}\left(\dr A_{\theta}^{a}\right)^2 
		+ V(\ph{10}) \right\} } \\
  &=& 2\pi\int_{\rs}^{\rl} r dr \left\{ \frac{\ets}{2}\right.&h'^2 + 
    \frac{\ets n^2 k^2}{2r^2}[1-b]^2 h^2 \\ 
& & &\left.{}+ \frac{5n^2 k^2}{4g^2}
\left(\frac{b'}{r}\right)^2 
+ \lambda_{2}\frac{\eta_2^4}{4}(h^2 - 1)^2 \right\} \\
  &=& \multicolumn{2}{l}{ 2\pi\ets \left\{\frac{|nk|}{4}+
\frac{n^{2}k^{2}}{2} \left(\frac{1}{2|nk|} - \frac{2}{2+2|nk|} 
+ \frac{1}{4+2|nk|}\right) \right\} } \\ 
& &\multicolumn{2}{l}{ {}+ \frac{5n^2 k^2}{2g^2 \rl^2} + 
\lambda_2 \frac{\eta_2^4 \rl^2}{4}
\left(\frac{1}{2} - \frac{2}{2|nk|+2} + \frac{1}{4|nk|+2}\right) } 
\end{array} \label{abE}\eequ
Thus, taking $\eta_2 \rl \sim 1$, the energy is of order $|nk|\ets$.
It is minimised by minimising $|k|$, subject to the restriction that
$m=n\left(k-\frac{1}{5}\right)$ is an integer. For $|n|<3$ this occurs at 
$k=\frac{1}{5}$, giving $m=0$. Thus for small $n$, although the
electroweak Higgs field has a string-like profile (it is approximately
zero at $r=0$), it does not wind around the string. This is in
agreement with earlier work by Alford and Wilczek \cite{Alford}. 
The electroweak symmetry is restored around it, and over a much larger
region than the other symmetry restoration, since $\rl \gg \rs$.
Putting $a(r)=1$ in (\ref{abEWphi}) and (\ref{abEWgauge}) will give
the \NO\ equations for a string with winding number $|nk|$, thus the
profile of $h(r)$ is not what would usually be expected of an abelian string.  

If $n$ is a multiple of 5 then the energy will be minimised by $k=0$
(as would be expected). In this case $h$ is constant for $r >\rs$.
Thus the approximate solution obtained with the above method is
$h(r)=C r^{|n|/5}$ for $r<\rs$ and $h(r)=1$ for $r>\rs$. This has
approximately the same form as $f(r)$. Although the electroweak
symmetry is still restored, it is over a much smaller region than in the
$k \neq 0$ cases. Throughout this region the electroweak Higgs field
winds. All of these results are completely independent of the choice
of $\ka$ and $\kt$.

Intuitively, it seems that the above solution has minimal energy.
However, it might be possible to get even lower energy by adding more or 
different $SU(5)$ gauge fields. If (for $|n|<3$) such gauge fields
caused the Higgs field to wind (in order to give a zero covariant
derivative at infinity), the energy would be higher, since it is
roughly proportional to winding number. Thus an alternative gauge
field ($\Rt$) would have to satisfy 
$(\ts + \Rt)\Hgg = 0$ to give $e^{i\theta[\tsnb +\Rt]}\Hgg = \Hgg$
for all $\theta$. So $\Rt\Hgg = -\frac{1}{5}\Hgg$ is needed, which has the
unique solution $\Rt = R$, and so the minimal energy solution has
been found.

\subsection{The Nonabelian $SU(2)$ Strings}

\subsubsection{$\lqg{}{}$ and $\Xg{}{}$ Strings}

Although at the GUT scale all the $SU(2)$ strings have the same
properties, this is not true at low temperatures, since they affect
\ph{10} differently. There are basically two different cases. The
generators corresponding to the \lqgpm\ and \Xgpm\ both annihilate the usual
vacuum expectation value of \ph{10}, and so the string gauge fields
have no effect on this symmetry breaking, and \ph{10} can be constant 
everywhere. There is still the possibility of symmetry restoration
from potential terms, which is due to variation of the \ph{126} and \ph{45} 
fields (see section \ref{NgaugeSymRest}), although this is less
significant. The other generators do have a non-trivial effect.

\subsubsection{$W$ Strings}

For the string with the \WRpm\ gauge fields, 
$\tsnb^{2} \Hgg = \frac{1}{4} \Hgg$, so 
$e^{2\pi in\tsnb}\pzero{10} = (-1)^n \pzero{10}$. Thus, in the presence of a
topological string (when $n$ is odd), as with the abelian string (when
$\frac{n}{5}$ in not an integer), it is necessary to replace \ts\ with
$\ts + R$. $R$ is chosen to give a single valued \ph{10}, while still
giving a vanishing covariant derivative at infinity. In order not to
alter the GUT string $R \Nr$ and $R\tsnb \Nr$ must both be zero. The
only possibility is $R = k_i \gen{\WL{i}}$, where $\gen{\WL{i}}$ is
the generator corresponding to the $\WL{i}$ boson. As with the GUT nonabelian
string, the Higgs field needs to be split up into different
eigenvectors of $n^2 (R+\tsnb)^2$. Noting that $R^2 \Hgg=k^2 \Hgg$
and $R\tsnb \Hgg=\tsnb R\Hgg$, where $k^2=\sum_i k_i^2$, it
is found that for non-zero $k$, \ph{10} is made up of two such eigenvectors 
\bequ
\psi_{p,q} = \frac{1}{2}\Hgg \pm \frac{1}{k}R\tsnb\Hgg \label{nabEWeigen}
\eequ
which have eigenvalues $p,q = n\left|\frac{1}{2} \pm k\right|$. Both of
these must be integers, and so if $n$ is odd $2nk$ must be an odd
integer to ensure that \ph{10} is single valued.

If $n$ is even, it is sufficient for $nk$ to be an integer. In this
case $k=0$ is a possibility, and there is no need for extra gauge
fields. The field equations are solved using the ansatz of the form 
$\ph{10} = \frac{\eta_2}{\sqrt{2}}e^{in\tsnb\theta}\Hgg h(r)$. 

Without loss of generality $k>0$ and $p>q$ can be assumed when $n$ is odd.
This gives the following ansatz for \ph{10} and the modified gauge as
\bequa 
\ph{10} &=& \frac{\eta_2}{\sqrt{2}}
    e^{in(\tsnb + R)\theta}\left(\psi_p h_p (r) + \psi_q h_q (r)\right) \\
A_{\theta} &=& n\frac{2}{g}\left(\frac{a(r)}{r}\tsnb + \frac{b(r)}{r} R\right)
\hspace{1in} A_{\mu} = 0 \mbox{ \ otherwise} \label{EWWansatz}
\eequa
As with the other solutions,
$h_p(\infty)=h_q(\infty)=b(\infty)=1$, and $h_p(0)=b(0)=0$ to
give the correct asymptotic form of the solution and regularity at
$r=0$. If $q \neq 0$, $h_q(0)=0$ is needed too. 
The resulting field equations are
\bequ
h_p'' + \frac{h_p'}{r} -
 n^2\frac{\left[\frac{1}{2}(1-a)+k(1-b)\right]^2}{r^2}h_p =
\frac{2}{N_p \ets} \frac{\partial V}{\partial h_p} 
\eequ
\bequ
h_q'' + \frac{h_q'}{r} -
 n^2\frac{\left[\frac{1}{2}(1-a)-k(1-b)\right]^2}{r^2}h_q =
\frac{2}{N_q \ets} \frac{\partial V}{\partial h_q} 
\eequ
\bequ
b'' - \frac{b'}{r} = -g^2 \ets\left\{
   \left[(1-b)+\frac{1-a}{2k}\right]N_p h_p^2
 + \left[(1-b)-\frac{1-a}{2k}\right]N_q h_q^2 \right\}
\eequ
where $N_{p,q} = |\psi_{p,q}|^2$, and so $N_p + N_q = 1$.
Except for the extra gauge term, these resemble the usual nonabelian
vortex equations (\ref{nabequa}-\ref{nabequc}), like (\ref{abEWphi}) and
(\ref{abEWgauge}) resembled (\ref{abequa}) and (\ref{abequb}). As
with the abelian string the energy can be estimated and minimised with a trial
solution. When $r<\rl$ the effect of $b$ on $h_p$, $h_q$ can be neglected, and
$a$ can be taken to be 0 and 1 for $r$ less than and greater than
$\rs$ respectively. Thus for $r < \rs$, $h_p = C_p r^p$ and
$h_q = C_q r^q$. These contribute $r^{p+2}$ and $r^{2q+2}$ terms
to the gauge term $b$, which also has an arbitrary $r^2$ term. For
$r>\rs$, $h_{p,q} = A_{p,q}r^{nk} + B_{p,q}r^{-nk}$. Matching the
solutions at $r = \rs$ gives $B_{p,q} \sim A_{p,q} \rs^{2nk}$, thus
$B_{p,q}$ can be neglected. The gauge field then takes the form $b(r)
= F r^{2nk+2} + G r^2$, with $G$ arbitrary and $F$ related to
$B_p$ and $B_q$. Matching solutions at $r=\rl$ shows that the
$r^2$ term dominates. This suggests the trial solution
\bequ
h_p (r) = h_q (r) = \left\{ \begin{array}{c}
 \left(\frac{r}{\rl}\right)^{nk} \\  1 \end{array} \right.
b(r) = \left\{ \begin{array}{cl}
 \left(\frac{r}{\rl}\right)^2 & r < \rl \\
		  1 & r > \rl \end{array} \right. \label{nabtrial}
\eequ
where, as in (\ref{abtrial}), the solution is assumed to take its
asymptotic form for $r>\rl$, and the arbitrary constants are
determined by ensuring continuity. Substituting (\ref{EWWansatz}) into
the Lagrangian gives an additional contribution to the energy
(neglecting $r<\rs$) of 
\bequ \begin{array}{ll@{}l}
\mbox{Energy} =& 2\pi\int_{\rs}^{\rl} r dr &\left\{
   \frac{\ets}{2} (N_p h_p'^2 + N_q h_q'^2) +
   \frac{\ets n^2 k^2}{2r^2}[1-b]^2 (N_p h_p^2 + N_q h_q^2) 
\right. \\& &\left.{}+\frac{n^2 k^2}{2g^2}\left(\frac{b'}{r}\right)^2
+ V(h_p, h_q) 
						\right\} \label{nabE}
\end{array} \eequ
However, for $r > \rs$ the trial solution (\ref{nabtrial}) has
$h_p(r)=h_q(r)=h(r)$ where $h(r)$ is the Higgs part of the abelian
trial solution (\ref{abtrial}). Substituting this into (\ref{nabE})
reduces it to an expression almost identical to (\ref{abE}), the extra
abelian string energy. Thus, taking $\eta_2 \rl \sim 1$, the
electroweak $W$-string 
energy is of order $nk\ets$. For odd $n$, $2nk$ must be a positive odd
integer, so the minimal energy occurs when $k=\frac{1}{2n}$, and so
$p=\frac{1}{2}(n+1)$, $q=\frac{1}{2}(n-1)$. For even
$n$ the energy is minimised by putting $k=0$, and so there is no
symmetry restoration outside of the GUT string radius ($\rs$). Thus
only topological nonabelian $W$-strings will cause electroweak symmetry
restoration outside of the string core. This is in contrast to the
abelian case, in which symmetry is restored in the region $r>\rs$ for
odd and even $n$. 

The lowest energy string ($n=1$) will have $k=\frac{1}{2}$, although
this does not imply that the abelian string (with $|k|=\frac{1}{5}$)
has lower energy than the nonabelian string, since there is also the
contribution from the GUT part of the string, which is of order
$\eta_1^2 \gg \ets$, and is higher for the abelian string.
Despite the fact that only half the electroweak Higgs field winds
arround the string, in a similar way to the GUT Higgs field, all of the
Higgs field is approximately zero at the string's centre 
($h_0 \sim \sqrt{\rs/\rl}$ for $r < \rs$), and so the
electroweak symmetry is almost fully restored there. In contrast, the
symmetry breaking caused by the GUT Higgs field is only partially restored 
at the centre of the string, because $|\ph{126}|$ is about $\frac{1}{2}$
there. As with the abelian string the profile of \ph{10} is like that of
a string with a non-integer winding number, in this case $\frac{1}{2}$.

Unlike the abelian string there is a choice of extra gauge fields. For
$r>\rs$ they all give approximately the same contribution to the
energy, so it is necessary to consider the $r<\rs$ contribution to find
the precise minimal energy solution. Taking $k=\frac{1}{2n}$ and
matching solutions at $r=\rs$, gives 
$h_{p,q} = \sqrt{\frac{\rs}{\rl}}\left(\frac{r}{\rs}\right)^{p,q}$.
The extra energy contribution (ignoring potential and gauge terms) is then
\bequ
2\pi\frac{\ets}{2}\frac{\rs}{\rl}[N_p p + N_q q]
\eequ
Thus, since $p>q$, the energy will be minimised by minimising
$N_p$. $N_{p,q}$ can be found explicity in terms of the parameters
of the gauge fields. \ts\ is equal to 
$\frac{1}{2\sqrt{2}}(z\gen{\WR{+}} + z^{\ast}\gen{\WR{-}})$, for some
$z$ with $|z|=1$, and $R = k_i \gen{\WL{i}}$, with 
$\sqrt{\sum_i k_i^2}=k=\frac{1}{2n}$. Substituting these
expressions into (\ref{nabEWeigen}), and evaluating it gives
\bequ \begin{array}{r@{}l}
\psi_{p,q} = \frac{1}{2}[\ka \pm & \kt z^{\ast}(k_1+ik_2)/k]\Hg{0}
 + \frac{1}{2}[\kt \pm \ka z(k_1-ik_2)/k]\Hgt{0} 
 \\ & {}\pm \frac{1}{2}(z^{\ast}\kt \Hg{+} - z \ka \Hgt{+})k_3/k
\end{array} \label{psiexplicit} \eequ
and so $N_{p,q} = \frac{1}{2} \pm \ka\kt \mbox{Re} \{z(k_1-ik_2)/k\}$.
Working in a $U(1)_Q$ gauge in which $z$ is equal to 1,
$N_p = \frac{1}{2} + \ka\kt k_1/k$. Since $|k_1| \leq k$,
$N_p$ is minimised by putting $k_1=-k=-\frac{1}{2n}$,
which gives $k_2=k_3=0$. Thus the minimal energy solution has only
non-zero \WLpm\ gauge fields. Furthermore the ratio $\WL{2}/\WL{1}$is
equal to $\WR{2}/\WR{1}$. The energy difference between the various
choices of $k_i$ is only about $\frac{\rs}{\rl} \sim 10^{-13}$ of
the electroweak contribution, and so it is quite possible other
neglected effects could alter this choice. 

The above is generally true, although if $\ka$ or $\kt$ is
zero, the different choices of $R$ all give the same energy.
They are all gauge equivalent under $SU(2)$, but since this symmetry
is broken, there are several physically distinct minimal energy solutions.

An interesting case occurs when $\ka=\kt$. Then $\psi_p = 0$ and
$\psi_q = \Hgg$ giving
\bequ 
\ph{10} = \frac{\eta_2}{\sqrt{2}} h_q(r) \eith{n(\tsnb + R)}\Hgg
\eequ
so \ph{10} has just one winding number, instead of the usual two. In
this respect it resembles the equivalent abelian case, and will share
some of the properties of abelian strings (see section \ref{FermionZM}). 
There is a slight difference in that $(\tsnb+R)\Hgg$ is not
proportional to $\Hgg$. 

\subsubsection{$Y$ Strings}

The fourth type of nonabelian string has \ts\ proportional to an
appropriate linear combination of the generators of
the \Ygpm\ gauge fields. In this case $\tsnb^{2} \Hgg = \frac{1}{4}\Hgg$, 
and so the situation is similar to the $W$-string, and has a similar
solution. \ts\ will be equal to $\frac{1}{2\sqrt{2}}\left(c_i\gen{\Yg{i}{+}} 
+ c_i^{\ast}\gen{\Yg{i}{-}}\right)$, with $\sum_i |c_i|^{2}=1$.
To give a single valued \ph{10}, \ts\ must be
replaced with $\tsnb + R$, where $R$ does not alter the GUT
string, but still affects \ph{10}. The only such generators are
$c_i \gen{\Y{i}{+}}$, $c_i^{\ast}\gen{\Y{i}{-}}$, 
and a third generator made up of the $A$, $Z$, and gluon fields. 
Together they form an $SU(2)$ subgroup, with the third generator the
equivalent of $\gen{\WL{3}}$. Like the $W$-string,
the different choices of $R$ give an approximately equal increase in
energy at the electroweak scale. However, $\gen{\Y{i}{\pm}}$ have a non-trivial
effect on \ph{45}, so when the $SU(5)$ symmetry breaking is taken into
account, the choice of $R$ that minimises the energy will
be different to the corresponding choice for the $W$-string (see
section \ref{IntermedSymRest}). The $Y$-string is less physically
significant than the $W$-string, since although it gives a globally
defined charge, it gives multi-valued \WLpm\ fields. 

\subsection{Summary}

Electroweak symmetry is restored and electroweak string gauge fields
are present in the presence of the abelian and the $W$ and $Y$
nonabelian GUT strings. This generally occurs in a region around the
string whose size is inversely proportional to the electroweak Higgs
VEV, and is much bigger than the string core. If $n$ is a multiple
of 5 for the abelian, or 2 for the nonabelian strings, the region is
approximately the same as the string core, and there are no extra
string gauge fields.

It is also possible that \ph{10} will wind. For the abelian string its
winding number is the closest integer to $-\frac{1}{5}n$, and hence
zero for $n=1$. For the $W$ and $Y$ nonabelian strings it is $\frac{1}{2}n$ for
even $n$. For odd $n$, different parts of \ph{10} have different
winding numbers, a bit like the corresponding GUT string. They are
$\frac{1}{2}(n-1)$ and $\frac{1}{2}(n+1)$.
The remaining two nonabelian strings ($\lqg{}{}$ and $\Xg{}{}$) have no
effect on \ph{10} at all.

\section{Other Symmetry Restorations}
\label{OtherSymmRest}

\subsection{The Intermediate Symmetry Restoration}
\label{IntermedSymRest}

So far the effect of the string on the second Higgs field \ph{45} has
been neglected, because it is far less significant. \ph{45} is
in the adjoint representation, and so its covariant derivative takes
the form
$\Dmd\ph{45}=\partial_{\mu}\ph{45} -\frac{1}{2}ig\left[A_{\mu},\ph{45}\right]$.
The generator corresponding to the $S$ particle ($P$) commutes with
\ph{45}, so the gauge fields
of the abelian string will not stop \ph{45} from taking its usual
vacuum expectation value everywhere. Thus it has no effect on the electroweak 
symmetry breaking, and gives no additional contribution to the energy.
The other strings will give non-vanishing covariant derivatives at
infinity. This is avoided by allowing \ph{45} to wind like a string
\bequ 
\ph{45} = e^{in\theta \tsnb} \pzero{45}(r)e^{-in\theta \tsnb} \label{medansatz}
\eequ
where $\pzero{45}(\infty)$ is equal to the usual vacuum expectation value of
\ph{45}. Conveniently $e^{2\pi n i\tsnb}$ (for all choices of \ts) and 
\pzero{45} commute, so \ph{45} will be single valued for
all $n$, and no extra gauge terms are needed. As with the \ph{126} and
\ph{10} fields it is necessary to split \ph{45} up into eigenstates of
$\tsnb^2$. Thus, using the fact that $\pzero{45}(\infty)$ and
$\tsnb^2$ commute, and that $\tsnb^3 = \frac{1}{4}\tsnb$
\bequ \begin{array}{c}
\ph{45} = e^{in\theta \tsnb} \psi_1 s_1(r) e^{-in\theta \tsnb}
 				+ \psi_0 s_0(r) \\

\mbox{with } \hspace{.3in} \psi_1 = 2\tsnb^{2} \pzero{45}(\infty) 
- 2\tsnb \pzero{45}(\infty) \tsnb, \hspace{.2in}
\psi_0 = \pzero{45}(\infty)  - \psi_1
\end{array} \eequ
The field equations for $s_{0,1}(r)$ will be similar to (\ref{nabequa})
and (\ref{nabequb}), with the similar boundary conditions
$s_{0,1}(\infty)=1$ and $s_1(0)=0$. Since the gauge contribution
($1-a(r)$) vanishes for $r>\rs$, $\pzero{45}(r) = \pzero{45}(\infty)$
is a solution outside the string. Thus the region of symmetry
restoration will be of radius $\rs$ (order $|\ph{126}(\infty)|^{-1}$).
This is in contrast to the other Higgs fields, which restore symmetry
in regions of order the reciprocal of their own values at $r=\infty$.

This is true for the $\Xg{}{}$, $\lqg{}{}$ and $W$-strings. For higher
temperatures, at which the \ph{10} symmetry breaking has not occured,
it is also true for the $Y$-string. After the final symmetry breaking,
there is an additional contribution from the \Ypm\ fields. This
can be resolved by altering (\ref{medansatz}) to 
$\ph{45} = e^{in\theta (\tsnb+R)} \pzero{45}(r) e^{-in\theta (\tsnb+R)}$,
which is still single valued. If $R\pzero{45}(\infty) \neq 0$ this
will give a large increase in the energy. $R\pzero{45}(\infty)$ is only
equal to zero when $R$ involves just the $Y$-string equivalent of
$\gen{\WL{3}}$, so this possibility is most likely to occur. With the
$W$-string, $R\pzero{45}=0$ for all $R$, since \WLpm\ and \ph{45} commute,
so \ph{45} does not alter the energetically prefered choice of $R$. Thus as
previously stated, the \ph{45} symmetry breaking has little effect on
the string properties, except for the $Y$-string.

\subsection{The Minimal Energy Choice of \ts}

After the first symmetry breaking, there were only two gauge inequivalent
possible strings, and the nonabelian strings had the lowest
energy. After the second symmetry breaking, the $\Xg{}{}$ and
$Y$-strings, which are gauge equivalent under $SU(2)_L$ have
the lowest energy. Of the nonabelian strings, the $W$-string has
highest energy, with the $\lqg{}{}$-string having slightly less. The
final symmetry breaking gives an additional contribution to the
$Y$-string, so the most energetically favourable choice of string
generator is made up of $\gen{\Xg{i}{\pm}}$, and does not have any
effect on electroweak symmetry. Just because the other strings are not
energetically favourable does not mean that they will not form, but
just that they are less likely to form (but see section \ref{OtherGUTs}). 

\subsection{Non-gauge field symmetry restoration}
\label{NgaugeSymRest}

Even when a Higgs field is unaffected by a string's gauge fields, it
is still possible for symmetry restoration to occur via the potential
terms. This has previously been discussed for an abelian string in
\cite{Goodband}. For example, in the absence of strings, before
electroweak symmetry restoration occurs, so $\ph{10} = 0$, the
potential takes the form
\bequ
V(\ph{126},\ph{45}) =
 \lambda_1 \left(|\ph{126}|^2 - \frac{\eta_1^2}{2}\right)^2 
+ \lambda'\left(|\ph{45}|^2 - \frac{\eta'^2}{2}
 \left[\alpha+ \beta \frac{2}{\eta_1^2}|\ph{126}|^2\right] \right)^2
\eequ
where  $\eta'/\sqrt{2}$ is the usual VEV of \ph{45}, and $\alpha + \beta =
1$. This is minimised by setting \ph{126} and \ph{45} to their usual
VEVs. However, in the presence of an abelian string $|\ph{126}|$ is
proportional to $f(r)$, so, writing $|\ph{45}| = s(r)\eta'/\sqrt{2}$,
the potential becomes
\bequ
\lambda_1\frac{\eta_1^4}{4}(f^2-1)^2
+\lambda'\frac{\eta'^4}{4}(s^2-[\alpha+\beta f^2])^2
\eequ
which, for small $r$, is no longer minimised by $s = 1$, for $\beta \neq
0$. Thus, if the theory's parameters take appropriate values,
$|\ph{45}|$ will be lower than usual, or even zero at the string's
centre. Alternatively $|\ph{45}|$ could be higher than usual there. 

If symmetry restoration by this mechanism occurs at all, it will only
be in the region $r < \rs$, since $|\ph{126}|$, and hence
$V(\ph{126},\ph{45})$ take their usual values at higher $r$. Unlike
the corresponding symmetry restoration by gauge fields, \ph{45} will
never wind. When \ph{10} is not equal to its usual VEV, it could also
cause $|\ph{45}|$ to vary. In this case the symmetry restoration could
take place in the larger $r<\rl$ region.  

A similar situation can occur with \ph{10} in the presence of an
$\Xg{}{}$ or $\lqg{}{}$ string. In this case both $|\ph{45}|$ and
$|\ph{126}|$ are lower than usual for $r < \rs$. For nonabelian
strings the potential is more complicated since it involves $f_1$
and $f_0$ terms, as well as the corresponding \ph{45} terms. Even
inside the string, $f_{0}$ and $s_{0}$ are non-zero, so the variation
of the potential is likely to be less substantial than the abelian
case, and hence extra symmetry restoration is less likely to
occur. Even if it does, \ph{10} will always take its usual VEV outside
the string.

\section{Fermion Zero Modes}
\label{FermionZM}

In the presence of a string, non-trivial zero energy solutions to the
fermion equations of motion may exist. If such solutions
involve charged fermions moving along the string, they will correspond
to a superconducting current. The effects of such a current will be
observable at large distances, thus providing a possible method of
observing cosmic strings \cite{Witten}. The fermionic part of the
Lagrangian is
\bequ 
{\cal L}_{\mbox{\scriptsize fermions}}= \flb i\gamma^\mu \Dmd \fl 
- \frac{1}{2} i\gm{2} \flb \ph{10} \flc \\
- \frac{1}{2} i\gm{1} \flb \ph{126} \flc + (\mbox{h. c.})
\label{fermLag}
\eequ
where $\gamma^{\mu} = (1,\sigma^i)$ since $\fl$ is a two component
spinor. Varying $\flb$ in (\ref{fermLag}) gives
\bequ
 i\gamma^\mu \Dmd \fl 
- i\gm{2} \ph{10} \sigt \fl^\ast
- i\gm{1} \ph{126} \sigt \fl^\ast = 0 \label{fermEL}
\eequ 
Because of the symmetries of the system, it is possible to look for
solutions which are independent of $z$ and $t$. It is also possible to
separate out their $r$ and $\theta$ dependence. If such solutions are
found, it is easy to generalise them and allow $z$ and $t$ dependence.
The resulting equations are easily solved, and correspond to
superconducting currents \cite{Witten}. Thus the existence of
two-dimensional solutions implies that the string is superconducting. 
Following the approach of Jackiw and Rossi \cite{Jackiw}, $\fl$ can be
split into eigenvectors of $\sigma_3$, $\pU$ and $\pL$, and (\ref{fermEL}) 
becomes
\bequ
e^{i\theta}\left(\dr + \frac{i}{r}\dth
 + \frac{g}{2}A_{\theta}\right)\pU 
+ \gm{2} \ph{10} \pU^\ast + \gm{1} \ph{126} \pU^\ast = 0 \label{fermU}
\eequ
\bequ
e^{-i\theta}\left(\dr - \frac{i}{r}\dth
 - \frac{g}{2}A_{\theta}\right)\pL 
- \gm{2} \ph{10} \pL^\ast - \gm{1} \ph{126} \pL^\ast = 0 \label{fermL}
\eequ
Jackiw and Rossi considered the affect of an abelian string on a
system in which only one Higgs field was present. The presence of
additional Higgs fields complicates things. However \ph{126} has no
effect on most of the fermion fields, so the results of \cite{Jackiw}
can be applied in some cases.

It is not necessary to attempt to solve both (\ref{fermL}) and
(\ref{fermU}), since one can be transformed into the other. The
complex conjugate of (\ref{fermU}), with the substitutions 
$n \rightarrow -n$ and $\pU \rightarrow i\pL^{\ast}$, gives
(\ref{fermL}). Thus, although nonabelian strings with winding numbers
$n$ and $-n$ are gauge equivalent, it is convenient to consider them
both when solving (\ref{fermU}).

\subsection{Zero Modes for the Abelian String}

\subsubsection{High Temperature Neutrino Zero Modes}

At high temperatures \ph{10} is zero, and so with the exception of
$\nu^c$, none of the fermion fields are affected by Higgs fields.
For the conjugate neutrino (\ref{fermU}) becomes
\bequ
e^{i\theta}\left(\dr + \frac{i}{r}\dth
 + \frac{1}{2}n\frac{a(r)}{r} \right)\UU{\nu}^c
 + m_1 f(r) \eith{n} \UU{\nu}^{c\ast} = 0 \label{abferNeuGUT}
\eequ
where $m_1 = \gm{1}\eta_1/\sqrt{2}$. This equation has been
discussed in detail in \cite{Jackiw}. It has $n$ normalisable
solutions for $n>0$, and none otherwise. A similar equation is
obtained from (\ref{fermL}) (or by transforming
(\ref{abferNeuGUT})). This has suitable solutions if $n<0$, in which
case there are a total of $|n|$ of them. Thus conjugate neutrino zero modes
always exist at high temperatures in the presence of an abelian string, and
there are $|n|$ of them. For $r > \rs$ the solutions decrease
exponentially, so the zero modes are confined to the string core.
 
\subsubsection{High Temperature Non-Neutrino Zero Modes}

Although there is no Higgs field acting on the other fermion fields, it is
possible for zero modes to be generated by the string gauge fields, as
discussed by Stern and Yajnik \cite{Stern}. (\ref{fermU}) and
(\ref{fermL}) reduce to
\bequ
\left( \dr + \UL{\sigma} \left[\frac{i}{r}\dth + 
p_{\lambda}n\frac{a(r)}{r} \right] \right) \UL{\lambda} = 0 \label{Sternequ}
\eequ
where $\lambda = u_i, d_i^c, \mbox{etc.}$ (not $\nu^c$), and
$\UL{\sigma} = \pm 1$. $p_{\lambda}$ is the eigenvalue of the field
with respect to $\frac{P}{10}$, (so $p_{e^-} = -\frac{3}{10}$,
$p_{u_i^c} = \frac{1}{10}$, etc.). There are normalisable
solutions if $|np_{\lambda}| > 1$, all of which can be found analytically. 
The number of solutions is equal to the highest integer
that is less than $|np_{\lambda}|$. Thus $|n|$ must be at least
4 for any zero modes of this type to exist. If the only stable
strings have winding number 1, then only conjugate neutrino zero modes will
be present at high temperatures around an abelian string. 

\subsubsection{Low Temperature Non-Neutrino Zero Modes}

At lower temperatures \ph{10} is non-zero and so the situation is
different. With the exception of the neutrino fields, there is just
one Higgs field coupling to the fermions. Its two components have
winding numbers $m$ and $-m$, and (\ref{fermU}) reduces to 
\bequ
e^{i\theta}\left(\dr + \frac{i}{r}\dth + \frac{g}{2}A_\theta \right)
\UU{\lambda}^c + m_\lambda h(r) \epmith{m} \UU{\lambda}^\ast = 0
\label{abfermqc} \eequ \bequ
e^{i\theta}\left(\dr + \frac{i}{r}\dth + \frac{g}{2}A_\theta \right)
\UU{\lambda} + m_\lambda h(r) \epmith{m} \UU{\lambda}^{c\ast} = 0
\label{abfermq} \eequ
where $\lambda$ can be $d_i$, $u_i$, or $e^-$, and  
$m_u = \kt \gm{2} \frac{\eta_2}{4\sqrt{2}}$,
$m_e = m_d = \ka \gm{2} \frac{\eta_2}{4\sqrt{2}}$. The upper sign applies
for the $d_i$, $d_i^c$ and $e^\pm$ fields, since they couple
to the $\Hg{0}$ component of the Higgs field. The lower sign is taken
for the $u_i$ and $u_i^c$ fields, which couple to the $\Hgt{0}$
component. These equations are similar to those in \cite{Jackiw}, and
have been discussed in \cite{Stern}. For the down quark and electron
fields, they have $m$ normalisable solutions per particle type, if
$m>0$ (which only  occurs if $n<0$). The corresponding equations for
$\LL{\lambda}$ have $-m$ solutions for $m<0$. 
 For the up quark fields, there are $-m$ (or $m$ for
$\LL{u_i}$ equations) normalisable solutions per particle type for
$\UU{u_i}$ ($\LL{u_i}$) if $m<0$ ($m>0$). 
If $n$ is not a multiple of 5, the solutions decay exponentially
outside $r=\rl$. When $n$ is a multiple of 5 (in which case 
$m = -\frac{1}{5}n$), they decay outside $r=\rs$. Thus the zero modes are
confined to the region of symmetry restoration. 

The difference in
sign between up and down quarks in (\ref{abfermqc}) and
(\ref{abfermq}) has physical significance when time and $z$
dependence are added to the solutions. They then represent currents
flowing along the string, with the up quark current flowing in the
opposite direction to the down quark and electron currents.  

\subsubsection{Low Temperature Neutrino Zero Modes}

The situation is more complex for the neutrino fields since they are
affected by two Higgs fields at the same time. In this case
(\ref{fermU}) becomes
\bequ
e^{i\theta}\left(\dr + \frac{i}{r}\dth + \frac{n}{2}\frac{a(r)}{r} \right)
\UU{\nu}^c + m_\nu h(r) \emith{m} \UU{\nu}^\ast
+ m_1 f(r) \eith{n} \UU{\nu}^{c\ast} = 0 \label{abferNeuc}
\eequ \bequ
e^{i\theta}\left(\dr + \frac{i}{r}\dth  - \frac{3n}{10}\frac{a(r)}{r} 
- \left[m+\frac{n}{5}\right]\frac{b(r)}{r}  \right)
\UU{\nu} + m_\nu h(r) \emith{m} \UU{\nu}^{c\ast} = 0 \label{abferNeu}
\eequ
where $m_\nu = m_u$. Although Jackiw and Rossi did not consider this
case, it can be approached using a similar method to theirs.
The angular dependence can be removed with the substitutions
\bequ
\UU{\nu}^c = A \eith{l} + B^\ast \eith{(n-1-l)} \label{abferNeuThc}
\eequ \bequ
\UU{\nu} = C^\ast \emith{(m+1+l)} + D \emith{(m+n-l)} \label{abferNeuTh}
\eequ
(The case $2l = n-1$ will be considered later.) Ignoring the gauge
terms, the resulting four complex differential equations are
\bequ
\left(\dr - \frac{l}{r}\right) A + m_\nu h(r) C + m_1 f(r) B = 0 
\label{abNeua} \eequ \bequ
\left(\dr - \frac{n-1-l}{r}\right) B + m_\nu h(r) D + m_1 f(r) A = 0 
\label{abNeub} \eequ \bequ
\left(\dr + \frac{m+1+l}{r}\right) C + m_\nu h(r) A = 0 
\label{abNeuc} \eequ \bequ
\left(\dr + \frac{m+n-l}{r}\right) D + m_\nu h(r) B = 0 
\label{abNeud} \eequ
Splitting these equations into real and imaginary parts gives two
identical sets of four real equations, and so it is only necessary to look
for real solutions.
When $r$ is large, $h(r)$ and $f(r)$ are both approximately 1 and the
gauge terms can be neglected as they are of order
$\frac{1}{r}$. Eliminating $C$ and $D$, and then solving gives
\bequ
A+B \propto e^{\frac{1}{2}
  \left(-m_1 \pm \sqrt{m_1^2 + 4m_\nu^2}\right)}   
\mbox{\hspace{.2in}and\hspace{.2in}}		    A-B \propto e^{\frac{1}{2}
	\left(m_1 \pm \sqrt{m_1^2 + 4 m_\nu^2}\right)}   
\label{abNeuInfSol} \eequ
$C$ and $D$ are then proportional to $A$ and $B$ respectively.
There are four linearly independent solutions to these equations. They all
have exponential behaviour at $r=\infty$. Only two linear combinations are
normalisable there. 

Near the origin, to first order in $r$, $h(r)=Gr^{|m|}$, $f(r)=Fr^{|n|}$,
and the gauge terms can be dropped. Making these 
substitutions, the four real (or imaginary) solutions of
(\ref{abNeua}-\ref{abNeud}) near $r=0$ can be found to first order
\bequ \begin{array}{l@{\ \sim\ }l}
A & r^l, \spa r^{n+|n|-l}, \spa r^{|m|-m-l}, \spa r^{|m|-m+|n|-n+2+l} \\ 
B & r^{1+|n|+l}, \spa r^{n-1-l}, \spa 
				r^{|m|-m+|n|+1-l}, \spa r^{|m|-m-n+1+l} \\     
C & r^{|m|+1+l}, \spa r^{|m|+n+|n|+1-l}, \spa 
				r^{-m-1-l}, \spa r^{2|m|-m+|n|-n+3+l} \\
D & r^{|n|+|m|+2+l}, \spa r^{|m|+n-l}, \spa 
				r^{2|m|-m+|n|+2-l}, \spa r^{-m-n+l}  
\end{array} \label{abNeuZeroSol} \eequ
If $\varphi$ is solution of (\ref{abNeua}-\ref{abNeud}) for all $r$, which is
normalisable at $r=\infty$, then  it will have to match some
combination of the 2
normalisable solutions in (\ref{abNeuInfSol}) for large $r$. 
At $r=0$, $\varphi$ will be made up of a
combination of the solutions in (\ref{abNeuZeroSol}). So if $\varphi$
is to be normalisable everywhere, at least 3 of the solutions
(\ref{abNeuZeroSol}) must be well behaved at $r=0$.
Thus for each $l$ satisfying 3 of the inequalities $l \geq 0$, $l \leq -m-1$, 
$l \leq n-1$ and $l \geq n+m$, there will be one normalisable
solution. If $l$ satisfies all 4 there will be 2 solutions. Not all of
these solutions are independent since the real (or imaginary)
solutions for $l=l'$ and $l=n-1-l'$ are proportional.

For $l = \frac{1}{2}(n-1)$ the angular dependence of (\ref{abferNeuc})
and (\ref{abferNeu}) is removed with the substitutions
\bequ
\UU{\nu}^{c} = A \eith{l} \hspace{1in} \UU{\nu} = C^\ast \emith{(m+1+l)}     
\eequ
giving (after dropping gauge terms) the equations 
\bequ
\left(\dr - \frac{l}{r}\right) A + m_\nu h(r) C + m_1 f(r)A^{\ast} = 0
\eequ \bequ
\left(\dr + \frac{m+1+l}{r}\right) C + m_\nu h(r) A = 0
\eequ
These have $A$ and $C$ proportional to 
$e^{\left(m_1 \sigma \pm \sqrt{m_1^2 + 4m_\nu^2}\right)}$
for large $r$, with $\sigma = \pm 1$ depending on whether the real or imaginary
parts of $A$ and $C$ are being considered. For small $r$
\bequ \begin{array}{l@{\ \sim\ }l}
A & r^l, \spa r^{|m|-m-l} \\	C & r^{|m|+1+l}, \spa r^{-m-1-l} 
\end{array} \eequ
So in this case there is one real and one imaginary solution if 
$0 \leq l \leq -m-1$. 

This gives a grand total of $-2m$ ($-m$ real and $-m$ imaginary)
normalisable solutions if $m < 0$, and $0$ otherwise. Surprisingly this
does not depend on $n$. A similar approach can be applied to
(\ref{fermL}) to give $2m$ normalisable solutions, provided
$m>0$. Hence there are $2|m|$ possible neutrino zero modes after
electroweak symmetry breaking. As with the other particle zero modes,
they will be confined to the region of symmetry restoration.

For a topologically stable string $m=0$, so no neutrino zero modes form.
This is slightly surprising, since at higher temperatures, when \ph{10} is 
zero, the abelian string does have neutrino zero modes, and intuitively, since 
$\ph{10} \sim \frac{\eta_{2}}{\eta_{1}}\ph{126} \sim 10^{-13}\ph{126}$,
the situation would be the same for lower temperatures. 

\subsection{Zero Modes of the Nonabelian Strings}

There are two additional complications with nonabelian strings.
Firstly the particle states are not eigenstates of the string
generator, although this is easily solved by re-expressing the problem
in terms of gauge eigenstates. Secondly, there are effectively twice
as many Higgs fields, since each Higgs field has two parts with
different winding numbers and different profiles.

\subsubsection{High Temperature Neutrino Zero Modes}

At high temperatures the gauge fields are proportional to \ts. Since
$\nu^c$ is not an eigenstate of \ts, the equivalent of (\ref{abferNeuGUT})
is obtained by putting $\fl = (\vnc \pm 2\tsnb \vnc)\chi^{(\pm)}(r,\theta)$, 
which are eigenvectors of \ts. Their eigenvalues are $\pm \frac{1}{2}$.
Substituting this and (\ref{nabphis}) into (\ref{fermU}) gives
\bequ 
e^{i\theta}\left(\dr + \frac{i}{r}\dth \pm
\frac{na(r)}{2r}\right)\UU{\chi}^{(\pm)} 
+ m_1 \frac{1}{2} \left( f_0(r) \UU{\chi}^{(\mp)\ast} 
+ e^{\pm i n \theta} f_1(r) \UU{\chi}^{(\pm)\ast} \right) = 0 
\label{nabferNeuGUT} \eequ
(The equivalent equations from (\ref{fermL}) can be obtained by
complex conjugation.) These two equations bear some resemblance to
(\ref{abferNeuc}) and (\ref{abferNeu}), and the $\theta$ dependence
can be removed with the substitutions
\bequ
\UU{\chi}^{(+)} = A \eith{l} + B^\ast \eith{(n-1-l)}
\eequ \bequ
\UU{\chi}^{(-)} = C^\ast e^{-i(l+1)\theta} + D \eith{(l-n)}
\eequ
which are the same as (\ref{abferNeuThc}) and (\ref{abferNeuTh}), when
$m$ is equal to zero. The resulting $r$ dependent equations closely
resemble (\ref{abNeua}-\ref{abNeud}). Similarly they also have four independent
solutions, only two of which are well behaved at infinity, the rest
increase exponentially. For small $r$ they behave as badly as
(\ref{abNeuZeroSol}) with $m=0$. Thus there are no normalisable
solutions for any values of $n$. 

\subsubsection{High Temperature Non-Neutrino Zero Modes}

For the fermion fields that do not couple to \ph{126} it is possible
for zero modes to exist by the same mechanism as
(\ref{Sternequ}). However, unlike the abelian case, some fermion
fields are annihilated by \ts\, so $p_\lambda$ is effectively zero,
and they cannot have zero energy solutions for any value of $n$. For
instance, the $u_i$, $d_i$, $\nu$ and $e^-$ fields are all zero
eigenvectors of the string generator for the high temperature
$W$-string. Thus solutions can only occur for the conjugate fields, in
the presence of this type of string. Putting
$\fl = (\vl \pm 2\tsnb \vl)\chi^{(\pm)}(r,\theta)$, 
where $\tsnb\vl$ is not proportional to $\vnc$ or $\tsnb\vnc$, or equal
to zero, the nonabelian equivalent of (\ref{Sternequ}) is
\bequ
\left( \dr + \UL{\sigma} \left[\frac{i}{r}\dth \pm 
\frac{1}{2}n\frac{a(r)}{r} \right] \right) \UL{\chi}^{(\pm)} = 0 
\eequ
It has the solutions
\bequ
\UL{\chi}^{(\pm)} = r^l \exp \left( \UL{\sigma}\left\{il\theta 
\mp \frac{n}{2} \int_0^r ds\frac{a(s)}{s} \right\} \right)
\eequ
The solutions are normalisable if $0 \leq l < \pm \UL{\sigma} \frac{n}{2} - 1$.
Thus the total number of solutions, per number of particles (6
in this case), will be the largest integer below $\frac{1}{2}n$. In
order for any such solutions to exist $n$ must be at least 3, so
they do not occur for topologically stable strings. 

\subsubsection{Low Temperature $\lqg{}{}$ and $\Xg{}{}$ String Zero Modes}

At low temperatures \ph{10} is non-zero and couples to all the fermion
fields. When \ts\ is made up of generators of the $\lqg{}{}$ or
$\Xg{}{}$ fields, \ph{10} just takes its usual vacuum expectation
value. For the fermion fields that are not affected by \ph{126} there
is effectively no string and so no zero modes.
For the fields affected by \ph{126} the solutions of the field
equations will be at least as divergent as those of
(\ref{nabferNeuGUT}), so there will no normalisable solutions.

\subsubsection{Low Temperature $W$-String Non-Neutrino Zero Modes}
\label{LTWstrNnZM}

The neutrino and electron fields all couple to \ph{126} in the
presence of a $W$-string, while the quark fields are only affected by
\ph{10}. For a
topological $W$-string (so $n$ is odd), \ph{10} can be determined
using (\ref{psiexplicit}) and the comments after it
\bequ \begin{array}{llr@{}l}   
\ph{10} &=& \multicolumn{2}{l}{ \frac{\eta_2}{\sqrt{2}} e^{in\theta(\tsnb+R)} 
	\left\{ \frac{\ka - \kt}{2}(\Hg{0} - \Hgt{0}) h_p 
 	   + \frac{\ka + \kt}{2}(\Hg{0} + \Hgt{0}) h_q \right\} } \\
&=&\frac{\eta_2}{\sqrt{2}} \left[ \frac{\ka - \kt}{2} \right.&
\left\{ (\Hg{0} - \Hgt{0})\cos p\theta + i
(\Hgt{+} - \Hg{+})\sin p\theta \right\} h_p  \\ & &
{}+\frac{\ka + \kt}{2}&\left. \left\{ (\Hg{0} + \Hgt{0})\cos q\theta + i
(\Hgt{+} + \Hg{+})\sin q\theta \right\} h_q \right] 
\end{array} \label{nabstrHiggs} \eequ
$z$ has been gauge transformed to 1. Putting 
$\fl = a_i \sum_\pm (\vu{i} \pm \vd{i})\chi^{(\pm)} + 
(\vuc{i} \pm \vdc{i})\chi^{c(\pm)}$ (where $a_{i}$ are real) and the
expression for the Higgs field into (\ref{fermU}), and using the fact that 
$\vu{i} \pm \vd{i}$ and $\vuc{i} \pm \vdc{i}$ are eigenvectors of
$R$ and \ts, gives
\bequ \begin{array}{l}
e^{i\theta}\left(\dr + \frac{i}{r}\dth \pm
\frac{na(r)}{2r}\right)\UU{\chi}^{c(\pm)} \\ \hspace{1in}
+ \frac{\ka + \kt}{2} m_2 h_q \epmith{q} \UU{\chi}^{(\pm)\ast}
- \frac{\ka - \kt}{2} m_2 h_p \epmith{p} \UU{\chi}^{(\mp)\ast} = 0
\label{fermqc} \end{array} \eequ
\bequ \begin{array}{l}
e^{i\theta}\left(\dr + \frac{i}{r}\dth \mp
\frac{b(r)}{2r}\right)\UU{\chi}^{(\pm)} \\ \hspace{1in}
+ \frac{\ka + \kt}{2} m_2 h_q \epmith{q} \UU{\chi}^{c(\pm)\ast}
- \frac{\ka - \kt}{2} m_2 h_p \empith{p} \UU{\chi}^{c(\mp)\ast} = 0
\label{fermq} \end{array}\eequ
where $m_2 = \gm{2} \frac{\eta_2}{4\sqrt{2}}$. If $\ka \neq \kt$ the
angular dependence can then be separated with the surprisingly simple
substitutions
\bequ \begin{array}{ll}
\UU{\chi}^{(+)} = A \eith{l} & \UU{\chi}^{(-)} = B \eith{(p-q+l)} \\
\UU{\chi}^{c(+)} = C^\ast \emith{(1-q+l)} & 
\UU{\chi}^{c(-)} = D^\ast \emith{(1+p+l)}
\end{array} \eequ
where $l$ must be an integer. The resulting equations for $A,B,C,D$
have two sets of four solutions. One set is real, the other purely
imaginary. They both satisfy the same real equations. As with (\ref{abferNeuc})
and (\ref{abferNeu}) their behaviour for large and small $r$ can be
found. The solutions have exponential behaviour at large $r$, two are
divergent, two are normalisable. To first order for small $r$, four of
the terms present (one from each solution), are proportional to $r^l$,
 $r^{p-q+l}=r^{1+l}$, $r^{q-1-l}=r^{\frac{1}{2}(n-3)-l}$ and
$r^{-p-1-l}=r^{-\frac{1}{2}(n+3)-l}$.
In order to match up with some combination of the normalisable
large $r$ solutions, at least 3 of these must be well behaved at
$r=0$. This occurs when  $q-1 \geq l \geq 0$, in which case just 3 are
well behaved. Thus there are $2q = n-1$ (real and imaginary) normalisable
solutions to (\ref{fermU}), and (by complex conjugation)
$2q$ solutions of (\ref{fermL}).

If $\ka =\kt$, the $\epmith{p}$ terms in (\ref{fermqc}) and
(\ref{fermq}) are not present.
Apart from the gauge terms, these are practically the same as
(\ref{abfermqc}) and (\ref{abfermq}). They can be solved in the same
way, so there are $2q$ normalisable solutions. The corresponding
equations from (\ref{fermL}) also have $2q$ solutions. 

Since there are 3 linearly independent choices of $a_i (\vu{i} \pm \vd{i})$
there are a total of $12q$ different zero modes for the $W$-string
after electroweak symmetry breaking. The solutions are contained in
the $r<\rl$ region. Since $q=0$ for the energetically stable $n=1$
string, it has no fermion zero modes.

When $n$ is even (so the string is actually topologically equivalent
to the vacuum), \ph{10} is equal to
\bequ
\frac{\eta_2}{\sqrt{2}} \left\{ 
(\Hg{0}\ka + \Hgt{0}\kt)\cos \frac{n}{2}\theta 
+ i(\Hgt{+}\ka + \Hg{+}\kt) \sin \frac{n}{2}\theta \right\} h_{n/2}
\eequ
This is the same as (\ref{nabstrHiggs}) if $p$ and $q$ are both set equal
to $n/2$. Thus the results for the odd $n$ strings can be applied to even $n$
strings, and there are $12\frac{n}{2}$ normalisable solutions.

The mass terms considered in this theory are not of the same form as the ones
usually used in $SU(5)$ GUTs. Usually the VEV of \ph{10} 
consists of only one component in the ${\bf 5}_{-2}$ representation
(equivalent to putting $\ka=1$ and $\kt=0$). This gives masses to the
down quarks and electrons. The up quarks (and in $SO(10)$, the
neutrinos) get masses from the $P$-charge conjugate of \ph{10}, which
transforms under the $\bar{\bf 5}_2$ representation. The
fermionic part of the Lagrangian is thus taken as 
\bequ \begin{array}{r@{}l}
{\cal L}_{\mbox{\scriptsize fermions}} = \flb i&\gamma^\mu \Dmd \fl -
 \frac{1}{2}i \flb \left( \gma\ph{10}+ \gmat\pht \right) \flc \\
 &{}- \frac{1}{2}i \gm{1} \flb \ph{126} \flc + (\mbox{h. c.})
\end{array} \label{altfermLag} \eequ
where \pht\ is the $P$-charge conjugate of \ph{10}. If 
$\ph{10} = \phi_\alpha \Hg{\alpha}$, then \pht\ is equal to 
$\phi_\alpha^\ast \Hgt{\alpha}$. If $\gma$ and $\gmat$ are
suitably defined, (\ref{altfermLag}) will be equal to
(\ref{fermLag}) in the vacuum, or in the presence of an abelian
string, so the two theories will be equivalent. 

If a nonabelian string which affects \ph{10} is present, the two
theories are different. If it is a $W$-string, $R$ will be equal to
$k_i \gen{\WL{i}}$. Since $\kt=0$, the only restriction on $k_i$ is
$k=\frac{1}{2n}$. Choosing a gauge in which 
$\tsnb = \frac{1}{2\sqrt{2}}[\gen{\WR{+}} + \gen{\WR{-}}]$, and taking
$n=1$, $\gma = \gmat$, $k_1=1$ and $k_2=k_3=0$, gives (from
(\ref{psiexplicit})) 
\bequ
\ph{10} + \pht = \frac{\eta_2}{\sqrt{2}}(\Hg{0} + \Hgt{0})h_1 \cos \theta 
\eequ
which is proportional to the quark and electron mass terms.
This implies the fermion masses vary with $\theta$, and actually
vanish at $\theta = \pm \frac{\pi}{2}$. Moreover this is true for
all $r$. Other choices of $k_i$, $\gma$ and $\gmat$ will also have a
$\theta$ dependent mass. Clearly this is not physically credible, and
so the Lagrangian (\ref{altfermLag}), despite being the most obvious
generalisation of the $SU(5)$ GUT, is unrealistic. 

\subsubsection{Low Temperature $Y$-String Non-Neutrino Zero Modes}

The $Y$-string can be approached in a similar way to the $W$-string,
although the resulting equations are more complicated. Choosing a
gauge in which 
$\tsnb = \frac{1}{2\sqrt{2}}(\gen{\Yg{1}{+}} + \gen{\Yg{1}{-}})$, the
energetically favourable choice of $R$ is
$s\frac{1}{2n}\gen{\Y{1}{3}}$, with $s=\pm 1$. $\gen{\Y{1}{3}}$
is the generator that forms an $SU(2)$ subgroup with
$\gen{\Y{1}{\pm}}$. \ph{10} is then equal to
\bequ \begin{array}{l}
\frac{\eta_2}{2\sqrt{2}}\left[
\ka(\Hg{0}+s\Hgt{1})\eith{sp}  + \kt(\Hgt{0}+s\Hg{1})\emith{sp}\right]h_p
 \\ + \frac{\eta_2}{2\sqrt{2}}\left[
\ka(\Hg{0}-s\Hgt{1})\emith{sq} + \kt(\Hgt{0}-s\Hg{1})\eith{sq} \right]h_q
\end{array}\eequ
The problem is best expressed in terms of fermion eigenstates of $R$
and \ts. Putting $\fl = \vvp{u}v + s\vvp{d}w + \frac{1}{\sqrt{2}}(\vvcp{u}
+s\vvcp{d})\chi^{(+)} +\frac{1}{\sqrt{2}}(\vvcp{u}- s\vvcp{d})\chi^{(-)}$, 
where 
$\vv{u}' = a_1 \vu{1} + a_2 \vuc{2} + a_3 \vuc{3}$ and
$\vv{d}' = a_1 \vep + a_2 \vd{3} - a_3 \vd{2}$ with $a_i$
arbitrary and real, gives 
\bequ 
e^{i\theta}\left(\dr + \frac{i}{r}\dth - \frac{sb(r)}{2r}\right) \UU{v} 
+ \frac{m_2}{\sqrt{2}} \kt \left(h_p \emith{sp} \UU{\chi}^{(-)\ast} 
+ h_q \eith{sq} \UU{\chi}^{(+)\ast} \right) = 0
\eequ
\bequ 
e^{i\theta}\left(\dr + \frac{i}{r}\dth + \frac{sb(r)}{2r}\right) \UU{w} 
+ \frac{m_2}{\sqrt{2}} \ka \left(h_p \eith{sp} \UU{\chi}^{(+)\ast} 
- h_q \emith{sq} \UU{\chi}^{(-)\ast} \right) = 0
\eequ
\bequ
e^{i\theta}\left(\dr + \frac{i}{r}\dth 
				+ \frac{sna(r)}{2r}\right)\UU{\chi}^{(+)} 
+ \frac{m_2}{\sqrt{2}} \left(\ka h_p \eith{sp} \UU{w}^\ast 
+ \kt h_q \eith{sq} \UU{v}^\ast \right) = 0 
\eequ
\bequ
e^{i\theta}\left(\dr + \frac{i}{r}\dth 
				- \frac{sna(r)}{2r}\right)\UU{\chi}^{(-)} 
+ \frac{m_2}{\sqrt{2}} \left(\kt h_p \emith{sp} \UU{v}^\ast 
+ \ka h_q \emith{sq} \UU{w}^\ast \right) = 0
\eequ
These equations are similar to (\ref{fermqc}) and (\ref{fermq}). Their
$\theta$ dependence is removed with the substitutions 
\bequ \begin{array}{ll}
\UU{v} = A \eith{l} & \UU{w} = B \eith{(sq-sp+l)} \\
\UU{\chi}^{(+)} = C^\ast \emith{(1-sp+l)} & 
\UU{\chi}^{(-)} = D^\ast \emith{(sq+1+l)}
\end{array} \eequ
The resulting equations have the same form as the corresponding
$W$-string equations, and can be solved in the same way. This gives a
total of $12q = 6(n-1)$ (or $6n$ if $n$ even)
normalisable zero modes.

\subsubsection{Low Temperature $W$ and $Y$ String Neutrino Zero Modes}

For the fields affected by both \ph{126} and \ph{10} in the presence
of a $W$-string, the equations of motion are a combination of
(\ref{nabferNeuGUT}) and (\ref{fermq},\ref{fermq})
\bequ \begin{array}{l}
e^{i\theta}\left(\dr + \frac{i}{r}\dth \pm
\frac{na(r)}{2r}\right)\UU{\chi}^{c(\pm)} \\ \hspace{1in}
+ m_1 \frac{1}{2} \left( f_0(r) \UU{\chi}^{c(\mp)\ast} 
+ e^{\pm i n \theta} f_1(r) \UU{\chi}^{c(\pm)\ast} \right) \\ \hspace{1in}
+ \frac{\ka + \kt}{2} m_2 h_q \epmith{q} \UU{\chi}^{(\pm)\ast}
- \frac{\ka - \kt}{2} m_2 h_p \epmith{p} \UU{\chi}^{(\mp)\ast} = 0
\end{array} \eequ
\bequ \begin{array}{l}
e^{i\theta}\left(\dr + \frac{i}{r}\dth \mp
\frac{b(r)}{2r}\right)\UU{\chi}^{(\pm)} \\ \hspace{1in}
+ \frac{\ka + \kt}{2} m_2 h_q \epmith{q} \UU{\chi}^{c(\pm)\ast}
- \frac{\ka - \kt}{2} m_2 h_p \empith{p} \UU{\chi}^{c(\mp)\ast} = 0
\end{array}\eequ
The angular dependence is removed with the substitutions
\bequ \begin{array}{ll}
\UU{\chi}^{(+)} = A \eith{l} + E^\ast \emith{(p-q+1+l)} &
 \UU{\chi}^{(-)} = B \eith{(p-q+l)} + F^\ast \emith{(1+l)} \\
\UU{\chi}^{c(+)} = C^\ast \emith{(1-q+l)} + G \eith{(p+l)} & 
\UU{\chi}^{c(-)} = D^\ast \emith{(1+p+l)} + H \eith{(l-q)}
\end{array} \eequ
Only 4 of the 8 solutions are well behaved at large $r$. At small $r$,
no more than 4 solutions are well behaved for any choice of $l$. Thus
it is not possible to match up the different solutions to give one
which is normalisable everywhere. This is also true for the
$Y$-string, so neither of them have low temperature zero modes involving the
conjugate neutrino field.

\subsection{Summary}
The only fermion zero modes that form at high temperatures are $\nu^c$
zero modes around abelian strings (in which case there are $|n|$ of
them), or those that involve fermion fields that just couple to the
string gauge fields, and not \ph{126}. This latter type of zero mode
will only occur for higher $n$ strings ($|n| \geq 3$).

At low temperatures there are a total of $16|m|$ different
zero modes on an abelian string ($|m|$ for each particle type), where
$m$ is the winding number of \ph{10}. $m=0$ when $|n|<3$, so there are
no zero modes around topologically stable abelian strings, and hence
they can only be superconducting at low temperatures in the presence
of an unusual Higgs potential \cite{Witten}.

If $m \neq 0$, and $z$ and time dependence is added to the solutions,
they will correspond to superconducting fermion currents. The electron
and down quark currents will then flow in the opposite direction to
the neutrino and up quark currents. 

In the presence of a $\Xg{}{}$ or $\lqg{}{}$ nonabelian string there
are no zero modes at any temperature. The other types of nonabelian
string each have $12q$ zero modes ($q$ for each particle type not
coupling to \ph{126}), where $q$ is the winding number of the part of \ph{10}
which winds least, so $q = \frac{1}{2}n$ for even $n$, and 
$q = \frac{1}{2}(n-1)$ for odd $n$. For a minimal energy,
topologically stable string there are no fermion zero modes, although there
is still the possibility of superconductivity due to gauge boson zero
modes. Thus even the $X$ and $\lqg{}{}$-strings may be superconducting
\cite{ABC}. Indeed, it has recently been shown that such strings do
become current carrying by gauge boson condensation \cite{Yates}.

The supercurrents corresponding to any fermion zero modes present do
not consist of single particle types, as those around an abelian
string do. Instead they are made up of eigenstates of the string
generator. Also, unlike the abelian case, currents containing
each particle type flow in both directions along the string.

\section{Other Related Grand Unified Theories}
\label{OtherGUTs}

Although only one particular $SO(10)$ GUT has been discussed, many of
the results apply to different symmetry breakings. Any theory of the form 
\bequ
SO(10) \cdots \pharrow{126} \cdots 
 SU(3)_c \times SU(2)_L \times U(1)_Y \times Z_2
 \pharrow{10} SU(3)_c \times U(1)_Q \times Z_2 
\label{gensymbreak} \eequ
could have string solutions of the form (\ref{abansatz}) or
(\ref{nabansatz}), which would cause electroweak symmetry restoration
at low temperatures in the same way as (\ref{symbreak}). The form of the 
other Higgs fields will not make much difference, as long as they are single
valued in the presence of a string (like \ph{45}). If they are, it
will not be necessary to add extra gauge fields, and so \ph{10} will
have the same behaviour as in (\ref{symbreak}). The resulting
strings will have the same kind of zero modes as the (\ref{symbreak})
theory, provided none of the other Higgs fields couple to the
fermions. The only Higgs fields that can couple to fermions are those
which transform under a representation contained in the
${\bf 16} \times {\bf 16}$, since fermion mass terms transform as a
product of {\bf 16}s. The only such representations are {\bf 126},
{\bf 10}, and {\bf 120} (which is antisymmetric), so the results of
the previous section apply to a wide range of theories.

The Higgs fields which gain their VEVs after \ph{126} will
determine the most energetically favourable choice of string
generator, as \ph{45} did in (\ref{symbreak}). 
If a GUT of the form (\ref{gensymbreak}) has Higgs fields which take
non-zero VEVs before \ph{126}, the choice of \ts\ will be more
restricted. If a generator has already been broken, the formation
of the corresponding string will not occur.

One theory of the form (\ref{gensymbreak}) is
\bequa
SO(10) & \pharrow{A} & SU(5) \times U(1)_P 
\pharrow{45} SU(3)_c \times SU(2)_L \times U(1)_Y \times U(1)_P \\
& \pharrow{126} & SU(3)_c \times SU(2)_L \times U(1)_Y \times Z_2 \\
& \pharrow{10} & SU(3)_c \times U(1)_Q \times Z_2
\label{SUxUpsymbreak} \eequa
The \ph{A} Higgs field transforms under either the {\bf 45} or {\bf 210}
representation of $SO(10)$, and is an $SU(5)$ singlet. Unlike
(\ref{symbreak}), only abelian strings can form in this theory, since
the only generator that \ph{126} breaks is $P$. This means that
electroweak symmetry restoration will always occur in the presence of a string.
\ph{A} and \ph{45} will both take their usual VEVs in
the presence of such a string, so the only symmetries restored in
the string core will be $U(1)_P$, and the electroweak symmetry.
Another interesting feature of this theory is that strings can
form at energies close to the electroweak scale, and so
$\rs$ could be of similar size to $\rl$, although still smaller. This
string is a candidate for defect mediated electroweak baryogenesis
\cite{baryogenesis}. Switching the second and third symmetry breakings
also gives a theory with similar solutions. 

A different unifying gauge group (instead of $SO(10)$), with similar
properties to (\ref{SUxUpsymbreak}) is $SU(5) \times U(1)_P$. It was
suggested in \cite{Witten}, and has two independent gauge coupling
constants. Unlike (\ref{SUxUpsymbreak}), strings of all winding
numbers will be topologically stable, since $U(1)_P$ is broken to $Z$
instead of $Z_2$. They could still decay by splitting into several
strings with lower winding numbers. 
The field equations for the electroweak fields will
be the same as (\ref{abEWphi}) and (\ref{abEWgauge}), but with
$-\frac{1}{5}$ the ratio of the two couplings instead of just
$-\frac{1}{5}$. If the ratio is $\alpha$, then \ph{10}'s winding
number will be the nearest integer to $-\alpha \frac{n}{5}$ (with half
integers rounded towards zero), so if $|\alpha| > \frac{5}{2}$, \ph{10}
will always wind in the presence of a string.
Since $|m|$ will always be non-zero,
fermion zero modes will always be present. This also means that
neutrino zero modes can survive the electroweak phase transition

A theory which is substantially different from (\ref{symbreak}) starts
with the symmetry breaking 
$SO(10) \pharrow{54} SU(4) \times SU(2)^2 \times Z^C_2$. The
$Z^C_2$ symmetry is not the $Z_2$ symmetry in
(\ref{gensymbreak}). \ph{10} is not invariant under it, so it must be
broken during or before the electroweak symmetry breaking. This will
lead to formation of domain walls, and so such a theory will have
substantially different properties to (\ref{symbreak}), and is ruled
out cosmologically \cite{strings}.

Another type of theory closely related to (\ref{gensymbreak}) occurs
when \ph{126} is replaced by \ph{16}, where the usual VEV of \ph{16} is
proportional to $\Nr$. The gauge fields all gain masses in the
same way as the equivalent theory involving \ph{126}, but there
will be no discrete $Z_2$ symmetry, so there will be no topological
strings. However, solutions of the form
(\ref{abansatz},\ref{nabansatz}) can still form, although since 
$e^{2\pi i n \tsnb}$ will need to map $\Nr$ to $\Nr$ to give a
single valued \ph{16}, only solutions with even $n$ will occur. Of
course, if such strings are to be observed, they will need to be
stable, which will only happen for certain values of the theory's
parameters. Embedded defects similar to these have been discussed
previously \cite{Vachaspati,Nathan}.

Since \ph{16} does not couple to the fermions, $\gm{1}$ will be
zero in (\ref{fermLag}), and so the neutrinos will have the same kind
of zero modes as all the other particles (As would be the case if
\ph{126} were present, but $\gm{1}$ were zero). However such a theory
has left-handed neutrinos with significant masses, and so is not
compatible with the standard model (unless some other mechanism is
introduced to reduce the mass of the $\nu$ field).
 
Yet another set of related theories can be obtained from
(\ref{symbreak}) by choosing a different VEV of \ph{45}. Adding a
multiple of $P$ to it will not affect any of the $SU(5)$ symmetry
breaking since all the $SU(5)$ fields commute with it, thus it will
not alter which gauge bosons become superheavy. It will alter the
sizes of the masses of the $SO(10)$ fields. The most energetically
favoured choice of nonabelian string will be the one with the lowest
energy contribution at the \ph{45} symmetry breaking. This will be the
one whose string generator corresponds to gauge fields with the
lowest mass. So by choosing \ph{45} appropriately, any of the
nonabelian strings could become favourable. The $\Xg{}{}$ and $Y$
strings are gauge equivalent at this stage, but since the $\Xg{}{}$
string contributes nothing at the electroweak symmetry breaking, it
will always be more favourable than the $Y$ string. Thus any of the
$\Xg{}{}$, $\lqg{}{}$ or $W$-strings could be energetically
favourable. If it is the $W$-string, then it is most probable that
electroweak symmetry restoration will occur. The same sort of freedom
does not exist with \ph{126} and \ph{10}, since any such change will
give different fermion mass terms, and radically alter the theory.

\section{Conclusions}
\label{SOtenConc}

In this paper we have uncovered a very rich microstructure for
$SO(10)$ cosmic strings. In particular, we have found four nonabelian
strings as well as one abelian string. We have examined the effect of
the strings on the subsequent symmetry breakings and studied the zero
modes in detail. Our results are summarised in the table \vspace{.15in}\\
\begin{tabular}{cc|ccc|c|cc}
Gauge & Type & \multicolumn{3}{|c|}{Symmetry restoration}
& EW &\multicolumn{2}{|c}{Zero modes if} \\
field &  & $SO(10)$ & $SU(5)$ & EW & fields & GUT & EW \\
\hline
$S$ & abelian & yes & no & yes & $Z$ & $n\neq 0$ & $|n|\geq 3$ \\
\lqgpm & nonabelian & partial & partial & no & --- & $n\geq 3$ & never \\
\Xgpm & nonabelian & partial & partial & no & --- & $n\geq 3$ & never \\
\WRpm & nonabelian & partial & partial & yes & \WLpm & $n\geq 3$ & $n\geq 2$ \\
\Ygpm & nonabelian & partial & partial & yes & 
				$A,Z,$ gluons & $n\geq 3$ & $n\geq 2$
\end{tabular} \vspace{.1in}

It seems that electroweak symmetry
restoration by GUT strings is quite likely. The exact results are
dependent on the details of the theory and the choice of string
generator. For the $SO(10)$ theory considered, electroweak symmetry is
restored for the abelian string, and half the possible nonabelian
strings, although the most energetically favourable of these
does not restore electroweak symmetry. However, other closely related
$SO(10)$ GUTs have different minimal energy string solutions,
which will restore the symmetry, such as (\ref{SUxUpsymbreak}), or
(\ref{symbreak}) with a different choice of \ph{45}. Thus, our
results generalise to a range of theories. This is currently under
investigation. 

The size of the region of electroweak symmetry restoration for the
topologically stable ($n=1$) strings is determined by the electroweak
scale, and is much larger than the string core. For nonabelian strings, with
higher winding number, the region will be the same if they are
topologically equivalent to the $n=1$ string (i.e.\ odd $n$), and
restricted to the string core if they are topologically equivalent to
the vacuum (i.e.\ even $n$). There is no such distinction between
topological and non-topological abelian strings, which restore
symmetry in the larger region if the winding number is not a multiple of 5. 
Some of the $SU(5)$ symmetry is also restored by all of the nonabelian
strings, but not the abelian string. This is only within the string
core, irrespective of the string winding number, and since \ph{45} is
not zero there, the restoration is only partial.   

For any choice of \ts, ignoring possible potential driven
symmetry restoration, the GUT will not be fully restored at the
string's centre. All the nonabelian strings have non-zero (although
smaller than usual) \ph{126} and \ph{45} fields at their
centre, so the $SO(10)$ (apart from the electroweak fields) symmetry
is only partially restored inside the string. For the abelian string
\ph{126} is zero in the string core, but \ph{45} takes its usual
value, so with the exception of $U(1)_P$, most of the $SO(10)$
symmetry is broken. The resulting gauge boson masses are smaller, but
still superheavy. However, for the abelian and the $W$ and $Y$
nonabelian strings, there is almost full restoration of electroweak symmetry 
in a larger region than the string core.

Although the profile of the electroweak Higgs field obeys the same
boundary conditions as a string, its exact form has a closer
resemblance to a string with non-integer winding number. For the abelian
string the actual winding number of \ph{10} is less than that of the
GUT string (about $\frac{1}{5}$). The same is true for the nonabelian
string, which has the winding number (or numbers if $n$ is odd) of
\ph{10} about $\frac{1}{2}$ that of the string itself. 

For the abelian string it is the winding number of \ph{10} that
determines the existence of fermion zero modes after electroweak
symmetry breaking. The number of zero modes is 16 times its winding number, so
unfortunately there will be none for topologically stable strings, which
have $|n|=1$ and hence $m=0$. Neutrino zero modes can always exist at
high temperatures, but they do not survive the electroweak phase
transition (for $|n|=1$). This result is fairly general, as discussed
in section \ref{OtherGUTs}.

The cosmology of such strings is rather interesting. The existence of
neutrino zero modes at high temperatures enables the string to carry a
neutral current, and thus lead to the formation of vortons
\cite{vortona}. Normally, vortons formed at such high temperatures
result in the theory being ruled out cosmologically
\cite{vortona,vortonb}. However, in our case the vortons would cease
to be stable below the electroweak scale, and cannot be used to rule
out the theory \cite{vortonc}.

At high temperatures, it is also possible (for higher $n$) for zero
modes to form because of the string gauge fields. This can also occur
for the nonabelian strings, and for the non-conjugate neutrino fields
around abelian strings. However this effect is always overridden if
the fermion field couples to a non-zero Higgs field.

In the presence of a nonabelian string, different parts of \ph{10} can
have different winding numbers. In the cases considered it is the part
with the lowest winding number which determines the number of zero
modes. Its winding number is equal to $\frac{1}{2}n$ rounded down
to the nearest whole integer for $W$ and $Y$ strings, and 0 for the
other two types. There are then a total of 12 times this number of
possible fermion zero modes. The fields coupling to \ph{126} (part of
which has winding number 0) do not have such solutions. As with the
abelian string there are no zero energy fermion solutions for
topologically stable strings, and so fermion zero modes on strings are not as
common as would be expected. 

In our analysis we have only considered terms occuring in the tree
level Lagrangian. One-loop corrections are likely to induce couplings
between the nonabelian string field and the electroweak Higgs. This
may result in electroweak symmetry restoration around the $\Xg{}{}$
and $\lqg{}{}$ strings.
However, the electroweak Higgs field would not wind in this region, and
there would still be no fermion zero modes.

We wish to thank T. Kibble, W. Perkins and A. Yates for useful
discussions, and PPARC, Trinity College and the E.U. under the HCM
program (CHRX-CT94-0423) for financial support.

\appendix

\section{$SO(10)$ Grand Unified Theory}

Under $SO(10)$, all left-handed fermions transform under one
representation, and right-handed fermions transform under its
conjugate \cite{Ross}. It is
convienient to use just one representation. This can be achieved by
using the charge conjugates of the fermions 
($\psi_R^c = C\bar{\psi}_R^T$, $\psi_L^c = C\bar{\psi}_L^T$).
The charge conjugate of a right handed fermion transforms as a left handed
fermion, and vice versa. Thus $\fl = \psi_L + \psi^c_R$ is left
handed. For $SO(10)$ this definition is necessary, as well as
convienient. $\psi_L$ could be gauge transformed to $\psi_R^c$, so
any gauge invariant quantities will have to involve just $\fl$ and
$\fr$ (right-handed equivalent of $\fl$). However, $\fr$ is superfluous, 
since it is equal to $\sigt \fl^\ast = \flc$, so the theory can be described
entirely in terms of $\fl$. For the electron family, it can be written as 
\bequ
\fl^{(e)}= \left(u_1, u_2, u_3, \nu_e, d_1, d_2, d_3, e^-, d^c_1,
		d^c_2, d^c_3, e^+, u^c_1, u^c_2, u^c_3, \nu^c_e \right)^T
\eequ
where $d_i = d_{iL}$, $d^c_i= \sigt d_{iR}^\ast$, etc.\ so all
the fields are left handed. The other two families of fermions can be
described similarly. Adapting work by Rajpoot \cite{Rajpoot}, the
gauge fields can be expressed explicitly as $16\times 16$ matrices
\bequ 
A = \sqrt{2} \left( \begin{array}{llll}
 H & I_4 \WL{-} & M'_Y & M_X \\ 
 I_4 \WL{+} & H &  M'^\ast_X & M_Y \\ 
  M'^{\dag}_Y & M'^T_X & -H^\ast & I_4 \WR{-}\\
  M^{\dag}_X & M^{\dag}_Y & I_4 \WR{+} & -H^\ast 
 \end{array} \right) + \Lambda
\eequ 
where $I_4$ is the $4\times 4$ identity matrix, and
\bequ
M_X = \left( \begin{array}{llll} 0 & -\X{3}{-} & \X{2}{-} & -\Xg{1}{-} \\ 
\X{3}{-} & 0 & -\X{1}{-} &-\Xg{2}{-}\\ -\X{2}{-} & \X{1}{-} & 0 &-\Xg{3}{-} \\ 
  \Xg{1}{-} & \Xg{2}{-} & \Xg{3}{-} & 0 \end{array} \right)
\eequ \bequ
M_Y = \left( \begin{array}{llll} 0 & \Y{3}{-} & -\Y{2}{-} & -\Yg{1}{+} \\ 
-\Y{3}{-} & 0 & \Y{1}{-} &-\Yg{2}{+}\\ \Y{2}{-} & -\Y{1}{-} & 0 &-\Yg{3}{+} \\ 
  \Yg{1}{+} & \Yg{2}{+} & \Yg{3}{+} & 0 \end{array} \right)
\eequ 
$M'_X$ and $M'_Y$ are obtained by swapping \Xpm\ and \Ypm\ with
\Xgpm\ and \Ygpm\ in $M_X$ and $M_Y$. $H$ is defined as
\bequ
H = \left( \begin{array}{llll} \multicolumn{3}{c}{\mbox{\LARGE G}} &
 \begin{array}{l} \lqg{1}{-} \\ \lqg{2}{-} \\ \lqg{3}{-} \end{array} \\
 \lqg{1}{+} & \lqg{2}{+} & \lqg{3}{+} & 0 \end{array} \right)
\eequ 
$G$ is a $3\times 3$ matrix of containing the gluon fields, it is
hermitian, and so $H$ is too. The other fields are contained in the
diagonal matrix $\Lambda$
\bequ \begin{array}{lr@{}l}
\Lambda &= \mbox{diag}&\left(
(\Vs+\WL{3})_3, -3\Vs+\WL{3}, (\Vs-\WL{3})_3,-3\Vs-\WL{3}, \right.\\
& &\left.\hspace{.2in}
 (-\Vs+\WR{3})_3,3\Vs+\WR{3}, (-\Vs-\WR{3})_3,3\Vs-\WR{3} \right)\\
 &= \mbox{diag}&\left(
(s+2a+2z)_3, -3s+4z, (s-3z-a)_3,-3s-3a-z, \right.\\
& &\left.\hspace{.2in}	   (-3s+a-z)_3,s+3a-3z, (s-2a+2z)_3, 5s\right)
\end{array} \eequ
The subscripts indicate repeated values, and
\bequa
s &=& \frac{1}{5}\left( -\WR{3} + \sqrt{\frac{3}{2}}B' \right) \\
B &=& \sqrt{\frac{3}{5}}\WR{3} + \sqrt{\frac{2}{5}}B' \\
z &=& \frac{1}{4}\left(\WL{3} - \sqrt{\frac{3}{5}}B \right)\\
a &=& \frac{1}{4}\left(\WL{3} + \sqrt{\frac{5}{3}}B \right)
\eequa
$Z=\sqrt{10}z$ and $A=\sqrt{6}a$ are the unrenormalised electroweak
$Z^{0}$ boson and photon respectively. $S=-\sqrt{10}s$ is a
high energy $SO(10)$ boson. The generator $P$ is obtained by putting
$s=1$ and $a=z=0$ in the expression for $\Lambda$. The substitutions
$s=z=0$ and $a=\frac{1}{3}$ give the charge operator.

\subsection{Higgs Fields}

The electroweak Higgs field transforms under the ${\bf 5}_{-2}$ and
$\bar{\bf 5}_2$ representations of $SU(5)$. They are contained in
the {\bf 10} of $SO(10)$, and so the components of \ph{10} can be
expressed as symmetric products of spinors transforming under the {\bf
16} representation. Thus \ph{10} can be expressed as
$\phi_\alpha \Hg{\alpha} + \tilde{\phi}_\alpha \Hgt{\alpha}$, where
$\Hg{\alpha}$ ($\alpha = 0,+,1,2,3$), are the five components of 
${\bf 5}_{-2}$, and $\Hgt{\alpha}$ are the corresponding 
components of $\bar{\bf 5}_2$. $(\Hg{0},\Hg{+})$ and $(\Hgt{0},\Hgt{+})$ form
$SU(2)_L$ doublets, while $\Hg{i}$ and $\Hgt{i}$ ($i=1,2,3$) form an
$SU(3)_c$ triplet and antitriplet.
Expressing these components of {\bf 10} in terms of symmetric products
of {\bf 16}s gives   
\bequ \begin{array}{l@{\hspace{.3in}}l}
\Hg{0} = \frac{1}{4}[\sprod{\vd{j}}{\vdc{j}} + \sprod{\vep}{\ve}] &
\Hg{+} = \frac{1}{4}[\sprod{\vu{j}}{\vdc{j}} + \sprod{\vep}{\vn}] \\
\multicolumn{2}{l}{\Hg{i} = \frac{1}{4}[\eps{ijk}\sprod{\vdc{j}}{\vuc{k}} 
- \sprod{\vu{i}}{\ve} + \sprod{\vd{i}}{\vn} ]} \\
\Hgt{0} = \frac{1}{4}[\sprod{\vu{j}}{\vuc{j}} + \sprod{\vn}{\vnc}] & 
\Hgt{+} = \frac{1}{4}[\sprod{\vuc{j}}{\vd{j}} + \sprod{\ve}{\vnc}] \\
\multicolumn{2}{l}{\Hgt{i} = \frac{1}{4}[\eps{ijk}\sprod{\vu{j}}{\vd{k}}
- \sprod{\vdc{i}}{\vnc} + \sprod{\vuc{i}}{\vep}]}
\end{array} \eequ
where $\vn$ is the basis vector corresponding the $\nu$ field, etc.
The second Higgs field \ph{45} is in the ${\bf 24}_0$ component of
the {\bf 45} representation, and its usual vacuum expectation value is
proportional to the diagonal matrix
\bequ 
\mbox{diag}\left((1/3)_3, -1, (1/3)_3, -1, (2/3)_3, 2, (-4/3)_3, 0 \right) 
\eequ
In the absence of a string, \ph{126} is proportional to
$\sprod{\Nr}{\Nr}$, with $\Nr = \vnc$.

\subsection{Fermion Masses}

The masses of the fermions arise from Yukawa couplings to the Higgs
fields. These must of course be Lorentz and gauge invariant. The two
possible Lorentz invariant mass terms are Dirac masses 
($\bar{\psi}_L \psi_R+\bar{\psi}_R \psi_L$), and Majorana masses
($\bar{\psi}^c_L \psi_{L} +\bar{\psi}_L \psi^c_L +
 (\mbox{right-hand terms})$). Dirac masses transform as singlets under gauge
transformations, and so will not be gauge invariant when coupled to a
Higgs field. The Majorana masses transform as a product of {\bf 16}s,
and so can be coupled to similarly transforming Higgs fields, \ph{126}
and \ph{10}, but not \ph{45}. This gives the fermionic Lagrangian
(\ref{fermLag}), (considering only one family of fermions for
simplicity), with Yukawa couplings.

Of course, since (\ref{fermLag}) is invariant under $SO(10)$, it must
be invariant under $SU(5)$ as well. Thus \ph{126} can only couple to
$SU(5)$ singlets (i.e. products of the conjugate neutrino
field). The two components of \ph{10} couple to 
${\bf 5}_{-2}$ and $\bar{\bf 5}_2$ products of fermions. Under
$SU(5)$, the remaining fermions transform under ${\bf 10}_1$ and
$\bar{\bf 5}_{-3}$ representations. To find allowable mass terms,
products of these representations need to be expressed in terms of
irreducible representations. ${\bf 10}_1 \times {\bf 10}_1$ and 
${\bf 1}_5 \times \bar{\bf 5}_{-3}$ both contain $\bar{\bf 5}_2$s,
 and ${\bf 10}_1 \times \bar{\bf 5}_{-3}$ contains a ${\bf 5}_{-2}$. 
Thus, for the usual VEVs of the Higgs fields, the mass terms are
written in terms of particle fields as 
\bequ \begin{array}{lll}
\flb \Hg{0} \flc &=& \frac{1}{8}\left[ 
d_i^{\dag} \sigt d_i^{c\ast} + e^{-\dag} \sigt e^{+\ast} 
+ d_i^{c\dag} \sigt d_i^\ast + e^{+\dag} \sigt e^{-\ast} \right] \\ &=& 
 \frac{1}{4}\left[\bar{d}_{iL} d_{iR} + \bar{e}_L^- e_R^- \right]
\end{array} \eequ
and similarly for $\Hgt{0}$, so 
\bequ
\flb \phnb{10} \flc = 
\frac{\eta_2\ka}{4\sqrt{2}} \left[
 \bar{d}_{iL} d_{iR} + \bar{e}^-_L e^-_R \right]
+ \frac{\eta_{2}\kt}{4\sqrt{2}} \left[
 \bar{u}_{iL} u_{iR} + \bar{\nu}_L \nu_R \right]
\eequ 
\bequ
\flb \ph{126} \flc =
 \nu^{c\dag} \sigt \nu^{c\ast}\frac{\eta_1}{\sqrt{2}}
 = \nu^T_R \sigt \nu_R \frac{\eta_1}{\sqrt{2}}
\eequ
This model, unlike the standard model, has non-zero neutrino masses.
However, if the $\nu^T_R \sigt \nu_R$ term is much larger than the
$\bar{\nu}_L \nu_R$ term, the mass eigenstates will be
approximately $\nu_L$ and $\nu_R$, and have very small and very
large mass eigenvalues respectively, giving an almost massless
left-handed neutrino.

\end{document}